\documentclass[11pt]{article}
\usepackage{geometry}
\usepackage[dvipdfm]{graphicx}
\usepackage[FIGTOPCAP,nooneline]{subfigure}
\usepackage[T1]{fontenc}
\usepackage{lmodern}
\usepackage{bm}
\usepackage{amsmath, amsfonts, amssymb}
\usepackage{cite}
\usepackage{multirow}
\usepackage{afterpage, array, rotating}
\usepackage{url}
\usepackage{authblk}
\usepackage{color}
\usepackage{algorithm}
\usepackage[noend]{algorithmic}
\usepackage{bm}
\usepackage{mathtools}
\usepackage{ulem}
\usepackage{booktabs}
\usepackage{rotating}
\usepackage{longtable}
\usepackage{pdfpages}
\usepackage{float}

\newcolumntype{A}{>{\centering\arraybackslash}p{1.2cm}}
\newcolumntype{B}{>{\centering\arraybackslash}p{0.6cm}}

\title{
Researcher Population Pyramids: Tracking Demographic and Gender Trajectories Across Countries
}

\author[1]{Kazuki Nakajima}
\author[2, 3]{Takayuki Mizuno}

\affil[1]{Graduate School of Systems Design, Tokyo Metropolitan University, Japan.}
\affil[2]{Information and Society Research Division, National Institute of Informatics, Japan.}
\affil[3]{Joint Support-Center for Data Science Research, Research Organization of Information and Systems, Japan}

\begin{document}
\date{}
\maketitle

\begin{abstract}
The sustainability of the academic ecosystem relies on researcher demographics and gender balance, yet assessing these dynamics in a timely manner for policy is challenging. 
Here, we propose a researcher population pyramid framework for tracking demographic and gender trajectories across countries using publication data. 
We provide a timely snapshot of historical and present demographics and gender balance across 58 countries, revealing three contrasting patterns among research systems: Emerging systems (e.g., Arab countries) exhibit high researcher inflows with widening gender gaps in cumulative productivity; Mature systems (e.g., the United States) show modest inflows with narrowing gender gaps; and Rigid systems (e.g., Japan) lag in both. 
Furthermore, by simulating future scenarios, the framework makes potential trajectories visible. 
If 2023 demographic patterns persist, Arab countries' systems could resemble mature or even rigid ones by 2050. 
Our framework provides a robust diagnostic tool for policymakers worldwide to foster sustainable talent pipelines and gender equality in academia.
\end{abstract}

\section*{Significance Statement}
We introduce a researcher population pyramid framework for tracking cross-national demographics and gender balance. Applying this framework to 58 countries, we uncover three contrasting patterns in academic systems---which we term ``Emerging'', ``Mature'', and ``Rigid''---each facing unique challenges in researcher inflow and gender equity. For example, while many Arab countries show rapid growth in researcher inflow, they also exhibit widening gender gaps in cumulative productivity, signaling a critical policy window to foster equitable academic futures. This framework provides policymakers worldwide with a data-driven diagnostic tool to promote long-term sustainability and gender equality in academia.

{\flushleft{{\bf Keywords:} } science of science; academic sustainability; research ecosystem; research careers; gender equality}

\newpage

\section{Introduction}

The pursuit of global sustainability---addressing grand challenges such as climate change, public health crises, and equitable economic development---is inextricably linked to the vitality of the global research enterprise. 
Sustained innovation, driven by a robust and dynamic academic ecosystem, provides the essential knowledge, technologies, and solutions required to navigate these complex issues \cite{sachs2019}. 
However, the role of this ecosystem extends beyond immediate scientific discoveries; it is responsible for the long-term reproduction of research talent and the continuity of knowledge itself \cite{fortunato2018}. 
A disruption in this human capital pipeline, encompassing both the flow of researchers from early-career training to senior-level expertise and the diversity within that flow, undermines a society's capacity to address not only current problems but also future, unforeseen challenges \cite{hong2004, beninson2018, hofstra2020}.
Ensuring the sustainability of the academic ecosystem is not merely an internal concern for the research community but a fundamental prerequisite for the long-term well-being and resilience of global society as a whole \cite{oecd2021}.

The sustainability of the academic ecosystem, in turn, is fundamentally determined by the demographic composition of its researcher population \cite{gibbons1994, powell2004, zapp2022}. 
This composition across career stages---early-career, mid-career, and senior-career researchers---directly influences the quality and quantity of research outputs within countries.
Doctoral students and postdoctoral researchers constitute a significant part of the pipeline for future researchers, and individuals' productivity during postdoctoral periods correlates with subsequent career continuity \cite{duan2025}.
Importantly, across all career stages, researchers exhibit diverse productivity patterns \cite{way2017}, reflected in considerable variation in the timing of their most impactful research, which can occur at any stage of a research career \cite{sinatra2016}.
Mid-career and senior-career researchers also play critical mentoring roles, thus facilitating the effective transfer of scientific expertise to younger generations \cite{malmgren2010, ma2020}.
In addition to ensuring a balanced researcher population across career stages, achieving gender diversity in academia remains a persistent global challenge \cite{lariviere2013, west2013, bendels2018, holman2018, huang2020, dworkin2020, llorens2021, ni2021, teich2022}, as gender-diverse research environments foster broader research questions, varied perspectives, and enhanced scientific creativity \cite{fine2020, yang2022}.
Therefore, developing robust methods for the systematic analysis of researcher demographics and gender balance across countries has become a critical challenge for science policymakers and educational administrators worldwide.

Reconstructing researchers' publishing careers from bibliographic data has enabled broader international and longitudinal analyses of research systems and research careers, including productivity patterns \cite{sinatra2016, way2017}, collaboration patterns \cite{newman2001, wagner2005, li2019, kwiek2021}, research impacts \cite{waltman2016, wang2013}, structural imbalances \cite{lariviere2013, huang2020}, and many more \cite{zeng2017, clauset2017, fortunato2018}.
Quantifying researcher demographics across career stages, gender, and national contexts from bibliometric data, however, still poses a significant methodological challenge: reliably determining whether researchers' publishing careers are ongoing in specific years. 
This challenge stems from the diversity in publication intervals among researchers and technical barriers in identifying researchers' publications based on bibliographic data \cite{huang2020, yang2025, yang2025_2}.
This uncertainty, as assessments of total productivity and publishing-career length often rely on defining a career end \cite{huang2020}, necessitates a substantial time lag when evaluating researchers' career characteristics.
These delays are particularly problematic for rapidly evolving research systems, where timely insights are essential for informing effective research and educational policies.

Here, we propose a researcher population pyramid framework for mapping and comparing researcher demographics across publishing careers, gender, and national contexts.
Based on author and publication data from a large-scale bibliographic database, we deploy country- and gender-specific thresholds of inter-publication intervals to identify ``active'' authors (i.e., those whose publishing careers are ongoing) in a specific year.
Within this framework, we define the cumulative productivity of an active author as the number of their most recent, uninterrupted publications up to a given year.
Unlike the total productivity and publishing-career length, cumulative productivity can be measured without waiting for a publishing career to end.
We construct population pyramids by counting active authors according to their cumulative productivity in a given year by country and gender. 
This framework enables systematic international comparisons of researcher demographics and gender balance across time. 
For future contexts, we simulate potential trajectories based on plausible transition scenarios derived from observed trends.
Applying this diagnostic tool to 58 countries, we demonstrate its utility by revealing three contrasting evolutionary patterns---Emerging, Mature, and Rigid---exemplified by Arab countries, the United States, and Japan, respectively.

\section{Results}

\begin{figure*}[t]
\centering
\includegraphics[width=1.0\textwidth]{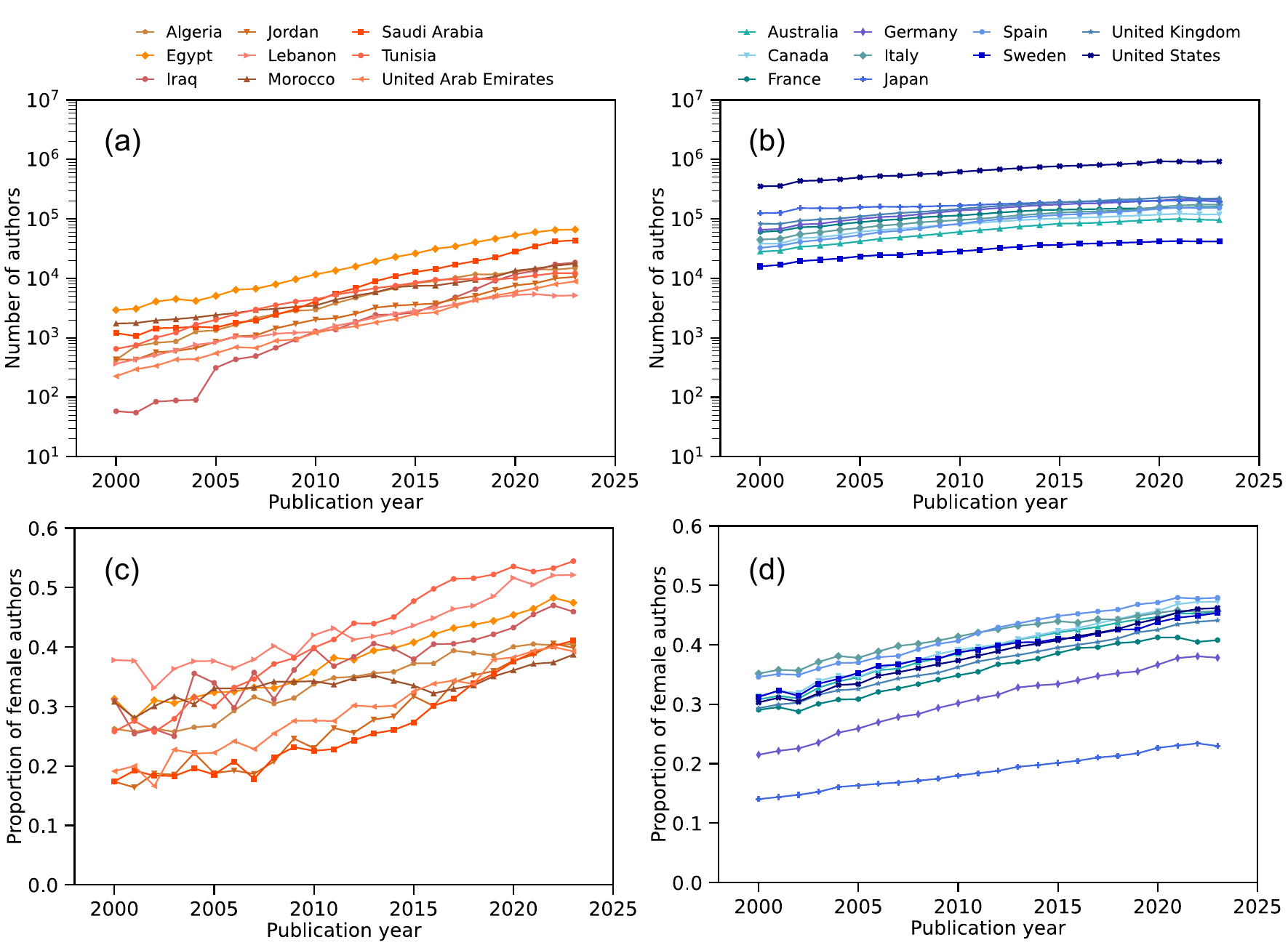}
\caption{Trends in researcher population and gender balance from 2000 to 2023.  
Panels (a) and (b): Annual number of unique authors who published at least one paper, affiliated with (a) Arab and (b) reference countries.
Plotted on a logarithmic scale. 
Panels (c) and (d): Annual proportion of female authors affiliated with (a) Arab and (b) reference countries.
}
\label{fig:1}
\end{figure*}

\subsection{Trends in Researcher Populations and Gender Balance}

We analyze 14,745,796 gender-assigned authors, derived from 151,905,632 publication records. 
These authors are affiliated with 58 countries for which Naive Bayes gender classifiers achieved high accuracy in two benchmarks (see Methods and Supplementary Sections S1 and S2).
For comparison, we focus on two country groups. 
The nine Arab countries are Algeria, Egypt, Iraq, Jordan, Lebanon, Morocco, Saudi Arabia, Tunisia, and the United Arab Emirates. 
The ten reference countries are Australia, Canada, France, Germany, Italy, Japan, Spain, Sweden, the United Kingdom, and the United States. 
We selected these countries to ensure a sufficiently large sample of gender-assigned authors (e.g., at least 9,000 per gender per country) and to represent regional diversity relevant for cross-national comparisons.
The comparison is primarily motivated by their contrasting developmental trajectories. 
In particular, Arab countries experienced significant socio-political transformations during and after the Arab Spring (2010--2012) \cite{tufekci2012, korotayev2014}.
Given these systemic changes and their notable increase in research output \cite{ibrahim2018, marzouqi2019, ahmad2021, unesco2021, almuhaidib2024}, we hypothesize that Arab countries are undergoing rapid shifts in researcher demographics and gender balance. 
By contrast, reference countries, with their established research infrastructures and stable funding mechanisms \cite{unesco2021}, serve as demographic benchmarks for comparison.

To quantify the long-term growth of researcher populations, we counted the number of unique gender-assigned authors who, for each year between 2000 and 2023, published at least one paper with their country of affiliation (Figs.~\ref{fig:1}(a) and \ref{fig:1}(b)). 
These annual author counts served as proxies for the size of national researcher populations. 
We observed roughly exponential growth in author counts for both country groups, with average annual growth rates (estimated from log-transformed data) higher in Arab countries than in reference countries (see Supplementary Section~S3.1). 
We also calculated the annual proportion of female authors among unique authors for each year (Figs.~\ref{fig:1}(c) and \ref{fig:1}(d)). Among Arab countries, Jordan, Saudi Arabia, and Tunisia showed the largest increases in female author proportions, while Morocco exhibited a relatively modest increase (Fig.~\ref{fig:1}(c); see Supplementary Section~S3.2). 
In contrast, reference countries generally displayed more gradual trends in female author proportions (Fig.~\ref{fig:1}(d)). 
Notably, Japan exhibited the lowest growth rate, consistent with persistent gender imbalances in Japanese academia \cite{nakajima2023}.

\subsection{Cumulative Productivity and Population Pyramids}

\begin{figure*}[p]
\centering
\includegraphics[width=1\textwidth]{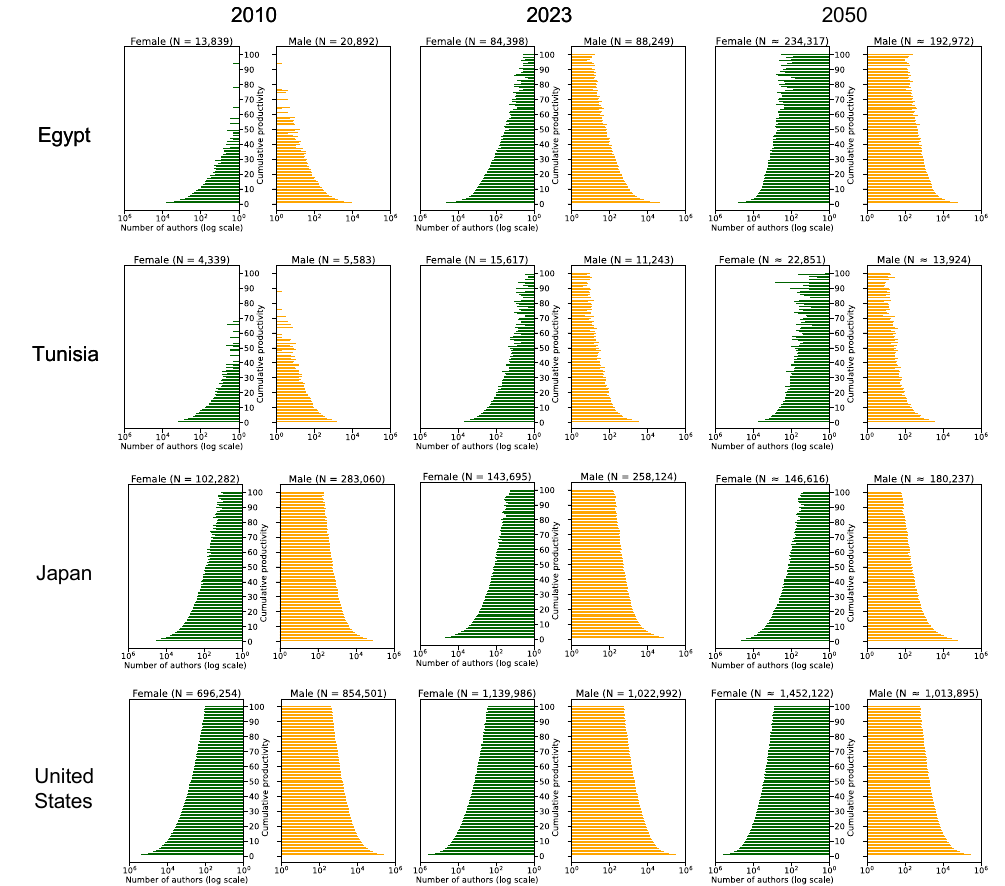}
\caption{Researcher population pyramids in 2010, 2023, and 2050 for Egypt, Tunisia, Japan, and the United States. The number of active authors for each gender is displayed on a logarithmic horizontal axis. Total active author counts ($N$) are provided for each panel. The 2050 pyramids and their corresponding counts ($\approx$) are projections based on 2023 trends.}
\label{fig:2}
\end{figure*}

To provide a more detailed understanding of researcher demographics, gender balance, publishing careers, and their evolution, we introduce a researcher population pyramid framework. 
This framework allows for the multi-dimensional visualization of researcher populations across publishing-career stages and genders. 
We construct these pyramids by counting active authors (i.e., those whose publishing careers are ongoing) in a given year by their cumulative productivity (i.e., the number of their most recent, uninterrupted publications up to that year; see Methods).
Unlike existing metrics that rely on a definitive career end, cumulative productivity can be measured without waiting for a publishing career to end.
We also simulate a plausible future demographic pattern by extending the 2023 pyramid forward (see Methods); our projections should be interpreted as scenarios based on 2023 trends.

Figure \ref{fig:2} illustrates the divergent demographic trajectories of researcher populations in four contrasting countries: Egypt, Tunisia, Japan, and the United States. 
To aid interpretation of these pyramids, we heuristically classify cumulative productivity into three descriptive stages: ``early-career'' (1--10 cumulative publications), ``mid-career'' (11--50 cumulative publications), and ``senior-career'' (51 or more cumulative publications). 
It should be noted that this labeling is for descriptive convenience only and is based solely on the length of an author's uninterrupted publication sequence, without accounting for research impact or differences across research disciplines.

\begin{figure*}[t]
\centering
\includegraphics[width=1.0\textwidth]{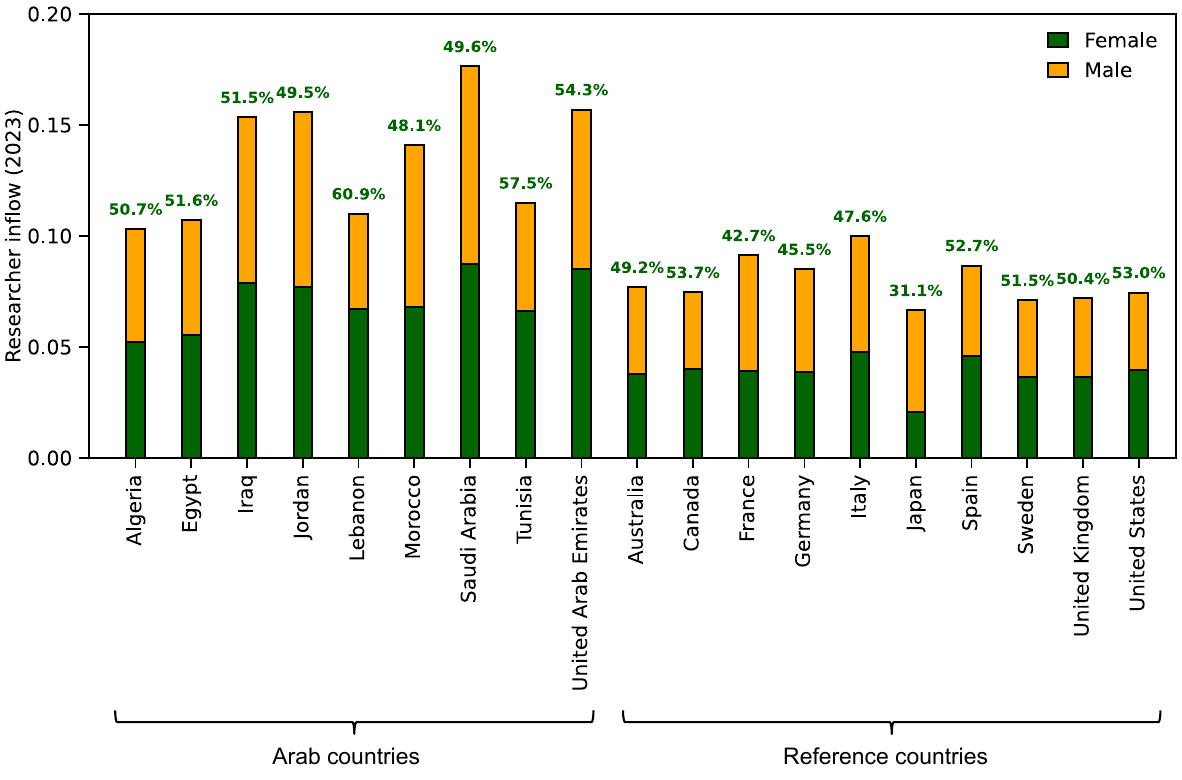}
\caption{Researcher inflow in 2023. Bars show the proportion of newly active authors in 2023 relative to all active authors in the same year. Green and orange segments represent female and male authors, respectively. Percentages above the bars indicate the female share among newly active authors.}
\label{fig:3}
\end{figure*}

In 2010, both Egypt and Tunisia exhibited relatively limited numbers of authors across publishing-career stages. 
Between 2010 and 2023, both countries experienced substantial expansion in their researcher populations, driven primarily by marked increases in early- and mid-career authors. 
Looking towards 2050, our projection scenario suggests substantial growth for Egypt, with both the number of authors (horizontal expansion) and cumulative productivity levels (vertical growth) continuing to expand. 
This suggests Egypt exemplifies a research system undergoing rapid demographic expansion. 
In contrast, Tunisia's 2050 projection suggests a more moderate overall growth characterized by a slowing inflow of early-career authors and an increasingly stable demographic profile. 
This pattern indicates Tunisia's research system is transitioning toward a more mature phase than Egypt's.

In 2010, Japan's research population pyramid showed a broad distribution of authors across publishing-career stages, but with a marked underrepresentation of women. 
This gender imbalance persisted through 2023, coupled with limited inflow at the early-career level. 
The 2050 projection indicates a continuation of this trend, with the pyramid remaining largely unchanged. 
These trends collectively depict a demographically rigid system, raising serious concerns about the long-term sustainability of Japan's research enterprise.

The United States' 2010 population pyramid showed a well-balanced distribution of researchers across all publishing-career stages, despite a moderate underrepresentation of women. 
This stable structure was maintained through 2023, with observable improvements in gender imbalance. 
Projections for 2050 suggest continued structural stability and only modest expansion. 
This trajectory---reflecting a research system that may be reaching saturation but appears demographically sustainable---presents a stark contrast to the dynamic growth observed in Egypt. 
Consequently, the United States exemplifies a mature, steady-state research system.

\begin{figure*}[t]
\centering
\includegraphics[width=1.0\textwidth]{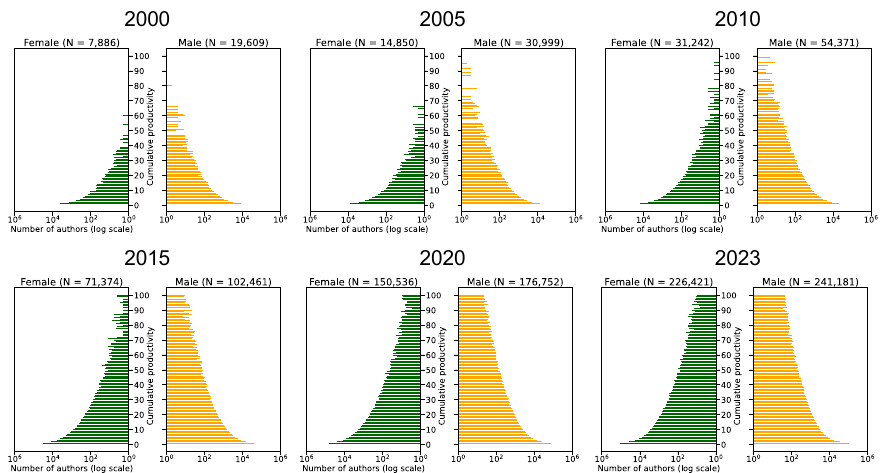}
\caption{Researcher population pyramids for the aggregate of the nine Arab countries in 2000, 2005, 2010, 2015, 2020, and 2023.}
\label{fig:4}
\end{figure*}

\begin{figure*}[t]
\centering
\includegraphics[width=1.0\textwidth]{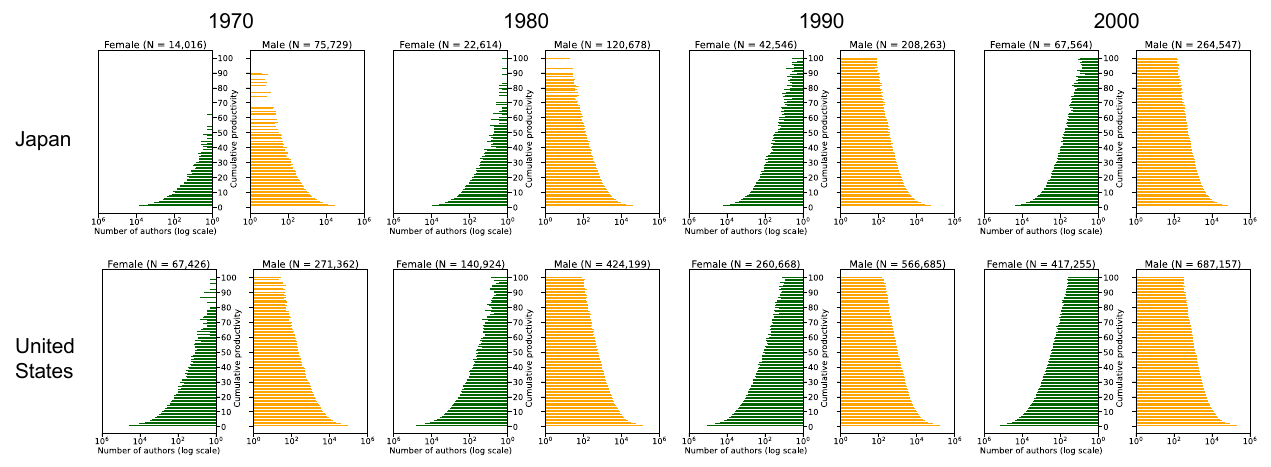}
\caption{Researcher population pyramids for Japan and the United States in 1970, 1980, 1990, and 2000.}
\label{fig:5}
\end{figure*}

\begin{figure*}[p]
\centering
\includegraphics[width=1.0\textwidth]{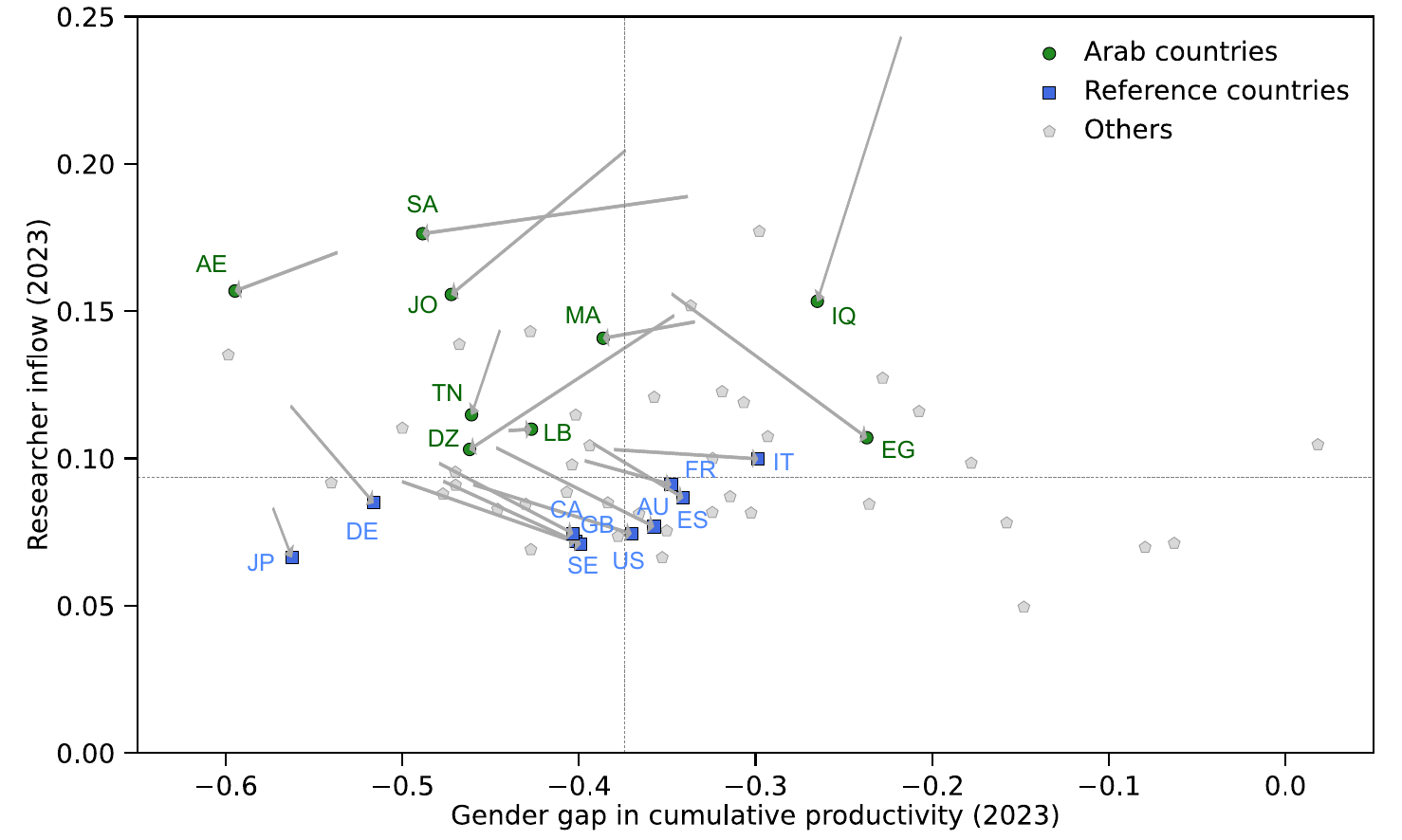}
\caption{
Scatter plot of the Arab, reference, and other countries positioned by researcher inflow (vertical axis) and the gender gap in cumulative productivity (horizontal axis) in 2023. 
Researcher inflow is the proportion of newly active authors among all active authors. 
The gender gap in cumulative productivity is defined as the mean cumulative productivity of female authors minus that of male authors, normalized by the male mean.
Country codes are as follows: AE (United Arab Emirates), AU (Australia), CA (Canada), DE (Germany), DZ (Algeria), EG (Egypt), ES (Spain), FR (France), GB (United Kingdom), IQ (Iraq), IT (Italy), JO (Jordan), JP (Japan), LB (Lebanon), MA (Morocco), SA (Saudi Arabia), SE (Sweden), TN (Tunisia), and US (United States).
Gray arrows indicate the trajectories of the Arab and reference countries from their coordinates in 2010 to their coordinates in 2023 in the same two-dimensional space. 
Vertical and horizontal dashed lines indicate the respective medians across the 58 countries. 
See Supplementary Section~S6 for detailed results for all 58 countries.
}
\label{fig:6}
\end{figure*}

Underlying the divergent shapes of the population pyramids, we found that countries vary greatly in their researcher inflow---defined as the proportion of newly active authors among all active authors affiliated with a country in a given year. 
For 2023, Arab countries showed substantially higher researcher inflow than reference countries (Fig.~\ref{fig:3}). 
Furthermore, the share of women among newly active authors achieves near gender parity (defined here as 45--55\% female representation) in most Arab countries and many reference countries. 
However, notable exceptions exist: Lebanon, Tunisia, and the United Arab Emirates exhibited slightly higher female shares than the defined parity range. 
In contrast, Japan consistently exhibited the lowest proportion of female newly active authors among all countries considered.

To provide a longitudinal perspective on the structural evolution associated with this researcher inflow, we examined the evolution of the aggregate researcher population pyramid, constructed by pooling all active authors affiliated with any of the nine Arab countries, from 2000 to 2023 (Fig.~\ref{fig:4}). 
In 2000, the pyramid exhibited a limited number of authors across all cumulative productivity levels. 
However, over the subsequent two decades, these countries experienced a marked horizontal widening of the early-career base, reflecting the substantial inflow of researchers.
By 2023, this sustained inflow had begun to translate into vertical growth, with an increasing number of authors reaching higher levels of cumulative productivity. 
This structural shift demonstrates how our framework can visualize the transition of the ``Emerging'' research system.

To contextualize the observed inflow dynamics and future projections for diverse research systems, we examine the long-term historical evolution of research populations in Japan and the United States from 1970 to 2000 (Fig.~\ref{fig:5}).
In 1970, Japan's pyramid featured a narrow base with very few senior female authors. 
By 1980, Japan's pyramid structure closely resembled that of the United States in 1970, suggesting Japan lagged about a decade behind the United States in its demographic development. 
However, the United States pyramid expanded rapidly thereafter; by the 1990s and 2000s, its shape mirrored that of Japan in 2023 (see also Fig.~\ref{fig:2}). 
This comparison implies that Japan now lags more than 20 years behind the United States in its population structure, indicating the development gap has widened. 
These retrospective pyramids provide a valuable reference point for interpreting the potential future trajectories of today's rapidly growing research systems, such as those in the Arab countries.

The preceding results lead us to position the 58 countries along two key dimensions: researcher inflow and gender gap in cumulative productivity (Fig.~\ref{fig:6}). 
First, all of these countries exhibit a negative gender gap in cumulative productivity (i.e., the mean for women is lower than for men), which is largely consistent with previous results on gender gaps in total productivity and publishing-career length across countries \cite{huang2020}. 
Second, the trajectories from 2010 to 2023 (gray arrows) suggest three broad, qualitative patterns of demographic evolution. 
For descriptive purposes, we heuristically label these patterns as follows:
\begin{itemize}
    \item ``Emerging'': Characterized by high researcher inflow and a widening gender gap in cumulative productivity (e.g., most Arab countries; arrows moving to the left).
    \item ``Mature'': Characterized by lower researcher inflow but a narrowing gender gap (e.g., the United States; arrows moving down and to the right).
    \item ``Rigid'': Characterized by low researcher inflow and a large, persistent gender gap (e.g., Japan; minimal arrow movement).
\end{itemize}
It is important to note that these labels are intended as a descriptive heuristic to facilitate interpretation. 
This qualitative classification is based on the observed positions and trajectories of countries within the specific two-dimensional space defined by the two metrics.
We also note that while most of the Arab countries analyzed exhibited a decrease in their researcher inflow from 2010 to 2023 (indicated by the downward direction of the gray arrows in Fig.~\ref{fig:6}), they consistently maintained researcher inflow above the global median (the horizontal dashed line). 
This trend suggests that while the initial rapid growth of their researcher populations has begun to stabilize as the total number of active authors increases, these systems continue to exhibit a substantially higher proportion of newly active authors relative to mature or rigid systems.

\begin{table*}[t]
\caption{Empirical and projected compound annual growth rates (CAGRs) of the number of active authors (2010--2050) and proportions of female authors among all active authors and senior-career authors (2023--2050) for selected countries. Country codes are as follows: AE (United Arab Emirates), AU (Australia), CA (Canada), DE (Germany), DZ (Algeria), EG (Egypt), ES (Spain), FR (France), GB (United Kingdom), IQ (Iraq), IT (Italy), JO (Jordan), JP (Japan), LB (Lebanon), MA (Morocco), SA (Saudi Arabia), SE (Sweden), TN (Tunisia), and US (United States).}
\centering
\small
\label{table:1}
\begin{tabular}{l  A A A A | B B B B | B B B B}
\\
& \multicolumn{4}{c|}{CAGR} & \multicolumn{8}{c}{Share of women (\%)} \\ 
 & 2010-- & 2023-- & 2030-- & 2040-- & \multicolumn{4}{c|}{All authors} & \multicolumn{4}{c}{Senior-career authors} \\
Country & 2023 & 2030 & 2040 & 2050 & 2023 & 2030 & 2040 & 2050 & 2023 & 2030 & 2040 & 2050 \\ \hline
AE & $0.166$ & $0.066$ & $0.031$ & $0.016$ & $51.5$ & $58.2$ & $62.3$ & $64.5$ & $17.6$ & $22.8$ & $36.8$ & $48.6$ \\
DZ & $0.133$ & $0.042$ & $0.024$ & $0.015$ & $47.2$ & $49.7$ & $51.3$ & $52.2$ & $13.2$ & $11.3$ & $13.5$ & $19.3$ \\
EG & $0.143$ & $0.058$ & $0.032$ & $0.019$ & $48.9$ & $51.5$ & $53.6$ & $54.8$ & $27.2$ & $38.7$ & $50.7$ & $57.1$ \\
IQ & $0.228$ & $0.091$ & $0.041$ & $0.019$ & $48.9$ & $51.1$ & $52.4$ & $53.0$ & $17.8$ & $29.7$ & $37.6$ & $42.6$ \\
JO & $0.136$ & $0.066$ & $0.029$ & $0.014$ & $43.6$ & $47.1$ & $48.8$ & $49.3$ & $18.6$ & $21.0$ & $28.1$ & $34.7$ \\
LB & $0.115$ & $0.044$ & $0.024$ & $0.014$ & $57.4$ & $59.5$ & $60.7$ & $60.8$ & $32.2$ & $33.3$ & $38.2$ & $42.7$ \\
MA & $0.133$ & $0.065$ & $0.032$ & $0.017$ & $41.4$ & $45.2$ & $46.7$ & $47.1$ & $21.8$ & $16.9$ & $23.6$ & $29.0$ \\
SA & $0.200$ & $0.076$ & $0.032$ & $0.016$ & $45.4$ & $49.9$ & $52.4$ & $53.3$ & $19.6$ & $31.4$ & $43.7$ & $48.9$ \\
TN & $0.081$ & $0.021$ & $0.011$ & $0.006$ & $58.1$ & $59.4$ & $61.0$ & $62.1$ & $32.7$ & $41.8$ & $54.2$ & $64.5$ \\
\hline
AU & $0.036$ & $0.005$ & $0.003$ & $0.002$ & $52.6$ & $54.1$ & $55.6$ & $56.6$ & $37.4$ & $42.2$ & $48.7$ & $54.1$ \\
CA & $0.029$ & $0.008$ & $0.005$ & $0.003$ & $55.1$ & $57.8$ & $60.2$ & $61.8$ & $35.2$ & $41.2$ & $49.6$ & $56.8$ \\
DE & $0.033$ & $0.007$ & $0.005$ & $0.003$ & $46.6$ & $49.4$ & $51.9$ & $53.6$ & $23.0$ & $28.9$ & $38.7$ & $47.8$ \\
ES & $0.048$ & $0.010$ & $0.005$ & $0.003$ & $53.7$ & $56.0$ & $58.0$ & $59.2$ & $38.3$ & $43.2$ & $50.1$ & $56.1$ \\
FR & $0.026$ & $0.007$ & $0.003$ & $0.002$ & $46.4$ & $46.2$ & $46.6$ & $47.0$ & $31.3$ & $36.8$ & $43.0$ & $48.2$ \\
GB & $0.031$ & $0.011$ & $0.007$ & $0.004$ & $52.7$ & $55.4$ & $58.0$ & $59.7$ & $33.4$ & $38.6$ & $45.8$ & $52.5$ \\
IT & $0.047$ & $0.018$ & $0.009$ & $0.005$ & $50.1$ & $51.3$ & $52.5$ & $53.2$ & $39.1$ & $43.9$ & $48.9$ & $51.6$ \\
JP & $0.012$ & $-0.015$ & $-0.007$ & $-0.003$ & $35.8$ & $39.9$ & $43.2$ & $44.9$ & $13.3$ & $17.2$ & $24.4$ & $32.9$ \\
SE & $0.029$ & $-0.011$ & $-0.007$ & $-0.004$ & $53.1$ & $56.6$ & $59.4$ & $60.8$ & $36.0$ & $41.2$ & $49.2$ & $56.2$ \\
US & $0.031$ & $0.007$ & $0.005$ & $0.003$ & $52.7$ & $55.2$ & $57.4$ & $58.9$ & $35.3$ & $41.7$ & $50.2$ & $57.3$ \\
\hline
\end{tabular}
\end{table*}

To assess the long-term sustainability of the research systems in these countries, we project their researcher population size and gender balance forward to 2050 based on their researcher population pyramids.
For population growth, we computed the compound annual growth rate (CAGR) of the number of active authors over four successive periods: 2010--2023, 2023--2030, 2030--2040, and 2040--2050 (see Supplementary Section S7 for CAGR definition). 
For gender balance, we examined the proportion of female authors among all active authors and among senior-career authors in 2023, 2030, 2040, and 2050.

Table \ref{table:1} reports empirical and projected CAGRs for active authors and the projected proportion of female authors across the selected countries. 
We observed a universal trend of projected decline in growth rates over time; this dynamic reflects our framework's assumption of stabilized numbers of newly active authors.
Arab countries exhibit greatly higher growth rates than reference countries for all the periods. 
In contrast, Japan and Sweden even approached negative growth.
Projections indicate divergent trajectories in the proportion of female authors among all authors and senior-career authors by 2050 across these countries.
The Arab countries, except for Jordan and Morocco, are projected to reach near gender parity (defined here as 45--55\% female representation) among all authors by 2050; Tunisia is even expected to slightly exceed this range.
However, the proportion of female authors among senior-career authors is projected to follow diverse trajectories across these Arab countries.
By contrast, reference countries show relatively stable progress; most are expected to reach near gender parity at both overall and senior levels by 2050, except for Germany and Japan.

\section{Discussion}

We examined researcher demographics across 58 countries through the lens of population pyramids, observing three broad patterns among research systems. 
We heuristically label these as follows: (i) ``Emerging'', exemplified by current Arab countries, which exhibit large inflows of newly active authors alongside widening gender gaps in cumulative productivity; (ii) ``Mature'', such as the United States, characterized by moderate inflows and gradually narrowing gender gaps; and (iii) ``Rigid'', such as Japan, which lag in both researcher inflows and progress on gender equality.

Our analysis shows that Arab countries exhibit notably higher researcher inflows than the reference countries. 
This pattern is partially consistent with previous reports on their expanding research outputs \cite{ibrahim2018, marzouqi2019, ahmad2021, unesco2021, almuhaidib2024}, increased scholarly attention \cite{asgari2025}, and enhanced funding investment \cite{curriealder2017}.
This rapid growth is likely driven by ambitious national strategies, such as Saudi Arabia's ``Vision 2030'' \cite{mohiuddin2023}, aiming to transition these countries toward knowledge-based economies.
The sustainability of this inflow hinges on future trends in higher education. 
UNESCO data show that the proportion of individuals graduating from higher education in Arab countries (25.7\% in 2023 \cite{uis_data}) remains lower than that in OECD states (41.8\% in 2023 \cite{uis_data}), suggesting substantial room for further expansion. 
Consequently, with continued policy investment in research infrastructure and the training of early-career researchers, this inflow of new talent can continue or even accelerate in Arab countries.
Nevertheless, this quantitative expansion also entails critical challenges. 
The widening gender gap in cumulative productivity identified in several Arab countries by our analysis is a prominent example. 
Ensuring the sustainability of this growth, therefore, requires policymakers to look beyond sheer numbers and address systemic factors, including the quality of education and the creation of robust supports for research careers.

Gender gaps in cumulative productivity were evident across most countries examined, yet the magnitude of these gaps varied substantially.
Among Arab countries, Egypt, Iraq, Tunisia, and the United Arab Emirates stand out for the rapid increase in the proportion of female authors since the early 2000s (see Fig.~\ref{fig:1}).
However, our analysis reveals that Tunisia and the United Arab Emirates exhibit larger gender gaps in cumulative productivity than Egypt and Iraq (see Fig.~\ref{fig:6}), which may be associated with greater attrition of women from academic careers at senior-career stages \cite{pell1996, spoon2023}.
This suggests that targeted research policies will be increasingly important for addressing these gender gaps in cumulative productivity in these countries. 
Such policies include correcting gender imbalances in resource allocation \cite{gasser2014}, implementing career support measures that accommodate childcare and family responsibilities \cite{morgan2021, spoon2023}, and actively promoting female researchers as role models \cite{clark2005}.
Among reference countries, Germany and Japan lag behind in both the long-term increase in the share of women and the narrowing of gender gaps in cumulative productivity. 
In Japan, these findings are consistent with the persistence of gender inequality across many societal domains \cite{wright1995} and with a recent study highlighting the structural persistence of gender imbalance in Japanese academia \cite{nakajima2023}.
In Germany, despite ranking 6th in the 2023 Gender Gap Index \cite{gender_gap_index_2023}, our findings align with the relatively low share of full-time equivalent female researchers \cite{unesco2021} and gender imbalance in professional promotions in the same country \cite{schroder2021}.
These results demonstrate that while national trajectories in gender inclusion vary widely, structural barriers to research career advancement for women persist even in countries with high societal aggregate gender equality.

Interpreting our projections within the assumption of stationarity in demographic transitions after 2023, Arab countries are likely to face a demographic turning point as researchers age and begin to retire. 
Supporting this view, the growth rates projected for Arab countries in the near future (2030--2040) mirror those observed in the reference countries' recent past (2010--2023; see Table \ref{table:1}), which suggests that Arab countries may reach this demographic inflection point by 2050. 
Projections of gender balance further illuminate these diverging trajectories (Table \ref{table:1}). 
Egypt, for instance, is on a trajectory toward gender parity, indicating convergence with a mature system. 
In contrast, Algeria, Jordan, and Morocco are expected to maintain persistent gender gaps among senior-career authors, suggesting a trajectory more aligned with rigid systems. 
Saudi Arabia and the United Arab Emirates show signs of gradual maturation but with slower progress toward gender parity among senior-career authors. 

Our analysis relies on the author disambiguation implemented in the OpenAlex database. 
We acknowledge the potential limitations of this method in accurately reconstructing publication records, particularly for authors from countries with high surname homogeneity and diverse transliteration practices \cite{tang2010, zhou2024, culbert2025}. 
Enhancing the reliability of our framework requires future work on two fronts.
First, it warrants systematic efforts to benchmark, evaluate, and improve the disambiguation accuracy for author names from various national contexts across different bibliographic databases. 
Second, while our approach makes a significant addition to previous work \cite{huang2020} by introducing country- and gender-specific thresholds for inter-publication intervals of authors, this could be further refined by incorporating discipline- or career-stage-specific thresholds. 
Such methodological refinements would open new avenues for research into the dynamics of academic talent pipelines, ultimately contributing to a more equitable and sustainable academic ecosystem.

Regarding the validity of using cumulative productivity as a proxy for career stages, we note that this metric is a functional proxy based on researchers' uninterrupted publication sequences rather than total productivity or publishing-career length. 
While total productivity and publishing-career length serve as measures of academic careers \cite{huang2020, yang2025, yang2025_2}, they often require identifying a definitive career end and thus entail a substantial time lag when evaluating researchers' career characteristics \cite{huang2020}. 
We employed cumulative productivity in our framework because it provides a timely snapshot of the active publishing workforce, serving as a data-driven diagnostic tool for monitoring research ecosystems.

While we heuristically labeled the bins of cumulative productivity as ``early-career'' (1--10 cumulative publications), ``mid-career'' (11--50), and ``senior-career'' (51 or more) to facilitate international comparisons, this classification may introduce potential biases in assessing researcher demographics and gender balance. 
First, despite the heterogeneity of publishing-career trajectories \cite{way2017}, cumulative productivity may conflate current publishing-activity levels with ``academic age'' (e.g., years since the first publication \cite{primack2009}). 
This, for instance, might categorize publishing-active researchers with a short academic age (e.g., those with 30 publications within five years of their first publication) as ``mid-career'', potentially overestimating their academic careers.
Second, publication practices vary significantly across disciplines. 
For example, average total productivity in a research career in biology (Male: 16.56, Female: 10.31) was substantially higher than in mathematics (Male: 7.13, Female: 5.55) in a previous analysis \cite{huang2020}. 
Our pyramids might reflect a complex interplay between national disciplinary composition and the underlying demographic distribution of publishing momentum.
Nevertheless, our additional analysis across four major research domains reveals that the qualitative characteristics of the Emerging, Mature, and Rigid research systems remain largely consistent within Egypt, Tunisia, Japan, and the United States (see Supplementary Section S8 for details). 
Third, the cumulative productivity metric is sensitive to career interruptions; a career break (e.g., due to parenthood \cite{morgan2021}) that exceeds the gender- and country-specific threshold resets the count to one in our framework. 
Although we defined these thresholds to accommodate diverse publication paces across countries and genders, the metric still prioritizes recent and continuous publishing momentum, potentially misclassifying experienced researchers with non-linear career trajectories as ``early-career''.

We acknowledge that these methodological choices and their potential biases might shape the interpretation of our results. 
First, the observed gender gap in the ``senior-career'' stage may reflect a disparity in publishing continuity. 
In ``Emerging'' systems, widening gaps in cumulative productivity may signify a ``publishing-continuity gap'', where female researchers---who disproportionately face career interruptions \cite{morgan2021}---could be assigned to lower cumulative-productivity bins. 
Consequently, our findings should be interpreted as highlighting the structural challenges women face in maintaining uninterrupted publication momentum. 
Second, the classification of research systems into ``Emerging'', ``Mature'', and ``Rigid'' systems reflects the aggregate dynamics of active publication momentum. 
For instance, a ``Rigid'' system like Japan's likely reflects not only low researcher inflow but also structural barriers or disciplinary compositions that prevent mid- and senior-career researchers from sustaining high publication momentum. 

Our approach has additional limitations.
First, the geographical scope of our analysis is constrained by the performance of name-based gender inference. 
We had to exclude several major research-producing countries, including China, India, and South Korea, where the classifier's accuracy was insufficient. 
This insufficient accuracy result is partially consistent with recent work demonstrating that the error rates of name-based gender inference are not uniformly distributed \cite{lockhart2023}.
Second, even among the countries included in our analysis, the gender assignment rate (i.e., the proportion of unique authors for whom a binary gender were inferred with high confidence) varies considerably (see Supplementary Table S7). 
For some countries, the rate is below 50\%; notably, for Iraq, a key country in our analysis of the Arab countries, the rate is 45.8\%. 
While we retained countries with lower rates in our analysis to maintain broad geographical coverage, their specific results should be interpreted carefully, as the subset of gender-assigned authors may not be representative of the entire researcher population in those countries.
Third, while we focus on the longitudinal visualization and classification of research systems in this study, our framework may be useful for quantitatively investigating the causal impact of specific sociopolitical events or policy interventions on researcher demographics. 
Future work could use this diagnostic tool to assess whether historical events (e.g., the Arab Spring in Arab countries \cite{ibrahim2018, asgari2025}) reshaped the generational profiles of academic systems or whether these systems remained resilient to such events.
Finally, our projections to 2050 are based on a strong stationarity assumption that the inflow in 2023 and transition patterns observed from 2022 to 2023 will remain constant in the future.
More sophisticated forecasting models that incorporate uncertainty, scenario variations (e.g., changes in researcher inflow), and structural breaks warrant future work.

Despite these limitations, our work highlights that understanding and developing the human capital of research has never been more critical in an era of global challenges. 
Our framework provides not just a retrospective snapshot but a dynamic diagnostic tool for policymakers to proactively shape more sustainable and equitable academic futures.

\section{Methods}

\subsection{Publication Data}

Our analysis is based on OpenAlex, a large-scale, open bibliographic database containing hundreds of millions of publication records with extensive metadata across multiple disciplines \cite{priem2022}. 
We used its September 27, 2024 snapshot, from which we extracted 151,905,632 publications classified as ``articles'' and published between 1950 and 2023. 
While our analysis focuses on this specific period (hereafter, $t_{\text{min}} = 1950$ and $t_{\text{max}} = 2023$), our framework is general and can be applied to any year range for which publication data are available.
For each paper, we extracted its publication date and, for each author of the paper, their ID, full name, and affiliations.
The author IDs are assigned by OpenAlex's proprietary author-name disambiguation algorithm.

We assigned a primary country (or countries) of affiliation to each author $u$ as follows. 
For each author $u$, we first compiled a list of all country codes (ISO 3166-1 alpha-2) from their affiliations. 
We then computed the frequency distribution of these countries and assigned the country (or countries) with the highest frequency as the primary affiliation(s) for $u$. 
If multiple countries tied for the highest frequency, we assigned all of them to $u$. 
We excluded any author for whom no country was assigned. 
Throughout our manuscript, for a given country $c$, we refer to an author whose assigned countries include $c$ as an ``author affiliated with country $c$.''

\subsection{Gender Assignment to Authors}

Bibliometric studies involving gender comparisons often require the assignment of gender to authors based on their first names (e.g., \cite{lariviere2013, dworkin2020, huang2020}).
Previous studies have shown that incorporating an author's country of affiliation improves the accuracy of such name-based gender inference \cite{huang2020, nakajima2023}.
Based on this, we trained the Complement Naive Bayes (CNB) classifier \cite{rennie2003, scikit-learn} that predicts the most likely binary gender and its associated probability, given an author's first name and country of affiliation.
We evaluated the classifier alongside an existing classifier \cite{vanbuskirk2023} using two large-scale, cross-country benchmarks: one derived from the Orbis executive database \cite{moodys_orbis}, and another from the World Gender Name Dictionary (WGND) \cite{wgnd_doc, wgnd_data} (see Supplementary Section~S1 for details).
For each of the 61 countries where the CNB classifier achieved sufficiently high accuracy, defined as an area under the receiver operating characteristic curve (AUC) of at least 0.8, on both benchmarks, we determined a country-specific confidence threshold $\theta_c$ for gender assignment (see Supplementary Table~S3 for details).
For this existing classifier \cite{vanbuskirk2023}, we also tuned a country-specific confidence threshold for gender assignment across the same countries (see Supplementary Table~S4 for details).
We excluded Bangladesh, China, India, and South Korea from our analysis because their AUC values fell below 0.8 in the WGND benchmark.

We thus began with the pool of 20,509,487 unique authors affiliated with one of the 61 countries. 
We first excluded 413,549 authors (2.0\% of the pool) for whom a single primary country could not be uniquely determined (i.e., multiple countries tied for the highest frequency). 
For each remaining author $u$ affiliated with country $c$, we extracted the first space-separated token from their full name (assumed to be the first name), and input it along with country $c$ into the CNB classifier to obtain an inferred gender $g$ and its associated probability $p_g$.
We assigned gender $g$ to $u$ if $p_g \geq \theta_c$; otherwise, we left the gender unassigned.
Finally, we excluded three countries (Lesotho, Papua New Guinea, and Somalia) from our analysis due to the small number of gender-assigned authors.
This process resulted in 14,745,796 gender-assigned authors affiliated with 58 countries. 
Supplementary Table~S7 reports the number of gender-assigned authors by country of affiliation.

We acknowledge that our classification is based on a binary conception of gender (female or male) inferred from first names. 
This approach, while common in large-scale bibliometric analyses \cite{lariviere2013, huang2020}, cannot account for individuals who identify as non-binary, transgender, or otherwise gender-diverse, and we recognize that it is an imperfect proxy for gender identity. 
Despite this limitation, we employ this approach as it remains one of the few feasible approaches for approximating gender demographics across countries.

\subsection{Inter-Publication Interval Threshold} \label{section:4.3}

Analyses of authors' publication records offer valuable insights into research career dynamics~\cite{sinatra2016, way2019, milojevic2018, huang2020, kwiek2023}.  
Key existing metrics, ``total productivity'', defined as the total number of publications in an entire career \cite{huang2020}, and ``publishing-career length'', defined by the duration between an author's first and last publications \cite{huang2020, yang2025, yang2025_2}, rely on a definitive ``last'' publication. 
For instance, Huang et al.~used a simple global rule: an author was considered to have ended their career by 2010 if no publications appeared after 2010 in records spanning 1900 to 2016~\cite{huang2020}.  
While easy to implement, this cutoff excluded all authors who continued publishing beyond 2010.

We refine this approach by classifying each author as ``active'' or ``inactive'' in a given year, based on the survival function of inter-publication intervals. 
These intervals---the time between consecutive publications---are measured in years by dividing the number of days by 365 and are estimated separately for each country and gender.
For each author who has published at least two papers, we extract their publication dates in chronological order to compute this sequence of intervals.
For each country $c$ and gender $g$, we pool all such intervals and estimate the corresponding survival function. 
We then determine a threshold, $\Delta_{\text{IPI},c,g}$, defined as the largest inter-publication interval (in years) at which the survival probability remains above 2\%. 

This 2\% threshold was chosen after our preliminary analysis (see Supplementary Table S10), as we determined this value strikes a good balance between maximizing the inclusion of authors with diverse publication paces and ensuring the threshold duration remains plausible.
A 1\% threshold, for instance, yielded excessively long inter-publication intervals (e.g., 15 years or longer) for several countries such as Jamaica.
Conversely, a 5\% threshold resulted in intervals of five years or less for most countries, which could prematurely classify authors as inactive in disciplines with slower publication cycles.

Using this threshold, we define an author as ``active'' in a given year $t$ if they published at least one paper between years $t - \Delta_{\text{IPI},c,g}$ and $t$. 
We consider authors who do not meet this criterion in year $t$ to be ``inactive.''

\subsection{Construction of Population Pyramids}

For a given country $c$, binary gender $g$, and year $t \in [t_{\text{min}}, t_{\text{max}}]$, we identify authors who satisfy a set of criteria extended from the methodology of Huang et al.~\cite{huang2020}. 
Specifically, we require that authors meet four conditions: 
(i) they have published at least one paper by year $t$; 
(ii) the duration between their first publication and the end of year $t$ does not exceed 40 years; 
(iii) their total number of publications by year $t$ is at most 500, a threshold corresponding to approximately the top 0.1\% of authors in our data;
and (iv) they are classified as ``active'' in year $t$, based on the survival function of inter-publication intervals described in Section \ref{section:4.3}.

To construct the population pyramid for a given country and year, we calculate the cumulative productivity as of that year for each active author affiliated with that country. 
This number, $k$, measures the continuity of an author's recent research activity, defined as the length of the author's most recent, uninterrupted sequence of publications.
We determine $k$ by chronologically scanning an author's entire publication record. 
We maintain a counter for the current continuous publication sequence. 
Starting with the first publication, this counter increases by one for each subsequent publication, as long as the time interval (in years) does not exceed the specified threshold ($\Delta_{\text{IPI},c,g}$). 
If an interval exceeds the threshold, the consecutive publication count ($k$) restarts at 1 with the next publication.
It is important to clarify that this procedure is not intended to suggest that a researcher's prior work becomes irrelevant after a prolonged publication interval.	
Rather, it serves as a practical strategy to mitigate potential errors in author disambiguation within the OpenAlex database.	
Since author disambiguation remains a technically challenging task~\cite{torvik2009, subramanian2021, culbert2025}, publication records belonging to different individuals may occasionally be conflated.	
By treating sufficiently long intervals in publication as breaks in the publication record, we aim to minimize the impact of such errors on our population pyramid estimates.
The final value of the counter after scanning all publications represents the author's cumulative productivity, $k$.
We provide a supplementary analysis of cumulative productivity including its correlation with total productivity and publishing-career length in Supplementary Section~S5.

To project the population pyramid for country $c$ in years beyond $t_{\text{max}}$, we propagate the population pyramid at year $t_{\text{max}}$ using the transition observed between year $t_{\text{max}} - 1$ and year $t_{\text{max}}$. 
We denote by $t^{\text{projection}}_{\text{max}}$ the end of the projection horizon beyond the observed period $[t_{\text{min}}, t_{\text{max}}]$; we set $t^{\text{projection}}_{\text{max}} = 2050$.
Let $n_{k,c,g,t}$ denote the number of authors of gender $g$ affiliated with country $c$ whose cumulative productivity is $k$ at year $t$.
For any $t_{\text{max}} + 1 \leq t \leq t^{\text{projection}}_{\text{max}}$, the projected population with cumulative productivity $k \geq 1$ evolves according to  
\begin{align}
n_{k,c,g,t} = a_{k,c,g,t} + \sum_{j \geq 1} n_{j,c,g,t-1}\,P_{j\to k,c,g,t-1},
\label{eq:1}
\end{align}
where 
\begin{itemize}
    \item $a_{k,c,g,t}$ is the number of 'newly active authors' at year $t$. These are authors who were inactive at year $t-1$ but become active at year $t$ with a cumulative productivity of $k$.
    \item $P_{j\to k,c,g,t-1}$ is the probability that an author with a cumulative productivity of $j$ in year $t-1$ moves to a cumulative productivity of $k$ in year $t$.
\end{itemize}
We obtain $n_{k, c, g, t_{\text{max}}}$ directly from the observed data.
Although $a_{k,c,g,t}$ and $P_{j\to k,c,g,t-1}$ are, in principle, time-dependent, we assume they remain fixed at their most recently observed values (namely, $a_{k,c,g,t_{\text{max}}}$ and $P_{j\to k,c,g,t_{\text{max}}-1}$) for every $t_{\text{max}} + 1 \leq t \leq t^{\text{projection}}_{\text{max}}$. 
Under this stationarity assumption, Eq. \eqref{eq:1} becomes  
\begin{align}
n_{k,c,g,t} \;=\; a_{k,c,g,t_{\text{max}}}\;+\;\sum_{j \geq 1}n_{j,c,g,t-1}\,P_{j\to k,c,g,t_{\text{max}}-1}
\label{eq:2}
\end{align}
for all $t_{\text{max}} + 1 \leq t \leq t^{\text{projection}}_{\text{max}}$.
We obtain every term on the right‐hand side from the historical publication record.  
Specifically, we count $a_{k,c,g,t_{\text{max}}}$ for every $k \geq 1$, and we recursively compute $n_{k,c,g,t}$ for all $k \ge 1$ and $t_{\text{max}} + 1 \leq t \leq t^{\text{projection}}_{\text{max}}$.  
We estimate the transition probability $P_{j\to k,c,g,t_{\text{max}}-1}$ as the fraction of authors with a cumulative productivity of $j$ in $t_{\text{max}}-1$ who moved to a cumulative productivity of $k$ in $t_{\text{max}}$.  
We use these empirical quantities in Eq.~\eqref{eq:2} and iterate forward to obtain projected population pyramids for years beyond $t_{\text{max}}$.
We emphasize that this analysis is based on a strong stationarity assumption regarding the inflow in $t_{\text{max}}$ and transition patterns observed from $t_{\text{max}} - 1$ to $t_{\text{max}}$.
This deterministic projection serves as a baseline scenario to illustrate potential long-term demographic trajectories if current conditions were to persist. 
This assumption is unlikely to hold over multi-decade horizons, as it does not account for cohort aging, policy shifts, or external shocks. 
The resulting projections should therefore be interpreted as stylized illustrations of current momentum.

\section*{Acknowledgments}

This manuscript was posted on a preprint: \url{ https://doi.org/10.48550/arXiv.2507.15500}.
We thank the anonymous reviewers for their constructive comments.
This work was supported in part by ROIS-DS-JOINT (004RP2025).
K.N.~thanks the financial support by the JST ASPIRE Grant Number JPMJAP2328 and by JSPS KAKENHI Grant Numbers JP24K21056 and JP25H01122.
T.M.~thanks the financial support by JSPS KAKENHI Grant Numbers JP23H00042 and JP25K01458.
T.M.~is grateful to Daiki Takeda from the Faculty of Law, The University of Tokyo, for insightful discussions on the classifier.

\section*{Declaration of Competing Interest}
The authors declare no competing interests.

\section*{Data Availability}

Publication and author data used in the analysis were extracted from the September 27, 2024, snapshot of OpenAlex \cite{priem2022}.
The World Gender Name Dictionary 2.0 Dataset was obtained from the WIPO Knowledge Repository \cite{wgnd_doc, wgnd_data}.
The April 2024 snapshot of Moody's Orbis database \cite{moodys_orbis} was obtained for a fee.
Access to Moody's Orbis database was licensed and cannot be shared publicly.
Information on how to request access is available at the following link: \url{https://www.moodys.com/web/en/us/capabilities/company-reference-data/orbis.html}.

\section*{Code Availability}
The code used for this manuscript is available at\\ \url{https://github.com/kazuibasou/researcher_population_pyramids}.

\newpage

\begin{center}
\vspace*{12pt}
{\Large Supplementary Information for:\\
\vspace{12pt}
Researcher Population Pyramids: Tracking Demographic and Gender Trajectories Across Countries}
\vspace{12pt} \\
\end{center}

\setcounter{figure}{0}
\setcounter{table}{0}
\setcounter{section}{0}

\renewcommand{\thesection}{S\arabic{section}}
\renewcommand{\thefigure}{S\arabic{figure}}
\renewcommand{\thetable}{S\arabic{table}}
\renewcommand{\theequation}{S\arabic{equation}}
\renewcommand{\thealgorithm}{S\arabic{algorithm}}

\begin{center}
\author{Kazuki Nakajima and Takayuki Mizuno}
\vspace{24pt} \\
\end{center}

\section{Gender Classification from Names}

This section details the construction and evaluation of the Naive Bayes classifier used to infer gender from author names. 
We describe the data sources for our training, validation, and test sets, the classifier, and the results of our benchmark evaluations.

\subsection{Data Sources and Benchmark Set Construction}

Our classifier was primarily trained and validated using data from Orbis, while WGND was used for additional tests of the model's generalizability. 
We detail the construction of benchmark datasets from both sources.

\subsubsection{Orbis Data}
We used the April 2024 snapshot of Moody's Orbis database \cite{moodys_orbis}. 
This database contains records for 69,020,795 executives, each providing their first name, binary gender, and a unique nationality. 
The nationalities in the dataset span 225 distinct ISO 3166-1 alpha-2 country codes.
We extracted executives from the Orbis database, grouping them by their nationality. 

We established specific data requirements for constructing training, validation, and test sets from Orbis data.
For each country included in our benchmarks, we aimed to create balanced validation and test sets, each comprising 450 female and 450 male executives. 
This specific sample size was selected primarily due to the limited data availability for Tunisia in the Orbis database, which recorded only 939 female executives.
We ensured that the training set for each gender was at least five times the size of the test set, necessitating a minimum of 2,250 individuals for training.
Because our analysis has a particular focus on Arab countries, we accommodated countries with varying levels of data availability. 
This strategy involved a specific measure for data-scarce Arab nations: we supplemented their training data by incorporating names from other Arab countries that possessed more extensive datasets.

Based on these criteria, we categorized and selected the countries according to the following three conditions:
\begin{enumerate}
\item The country belongs to the Arab League (see Table~\ref{table:s1}) and has at least 3,150 female and 3,150 male executives. This threshold directly corresponds to our requirement of a 450-person test set, a 450-person validation set, and at least a 2,250-person training set per gender.
\item The country belongs to the Arab League, has at least 900 female and 900 male executives (sufficient for a test set and a validation set), but has fewer than 3,150 individuals in at least one gender group.
\item The country does not belong to the Arab League and has at least 3,150 female and 3,150 male executives.
\end{enumerate}

This filtering process resulted in the identification of ten countries satisfying the first condition (i.e., Algeria, Bahrain, Egypt, Jordan, Kuwait, Lebanon, Morocco, Oman, Saudi Arabia, and the United Arab Emirates), four countries satisfying the second condition (i.e., Iraq, Somalia, Syria, and Tunisia), and 70 countries satisfying the third condition.
We excluded the countries that did not satisfy any of these conditions from the benchmark analysis.

For the 80 countries satisfying the first or third condition, we stratified and randomly partitioned 450 male and 450 female executives each to the test set and validation set, with the remaining individuals forming the country-specific training set.
For the four countries satisfying the second condition, we created the test and validation sets in the same manner. For their training set, however, we utilized a pooled Arab training set, aggregated from the training sets of all ten Arab countries that satisfied the first condition.
Finally, to address potential gender imbalance within the training sets, we downsampled the majority gender group to ensure the ratio of the minority group to the majority group was at least 0.2.

\subsubsection{World Gender Name Dictionary Data}

To test the generalizability of our classifier beyond the Orbis data, we used the World Gender Name Dictionary (WGND)~\cite{wgnd_doc, wgnd_data}.
The WGND ver.~2.0 contains 5,016,972 entries, each including a first name, binary gender, country, and the frequency of that name-gender combination within the country. 
These frequencies are primarily drawn from census and population registration data, though some sources reflect multi-year aggregations or alternative datasets~\cite{wgnd_doc}.

Among the 84 countries processed for the Orbis benchmark sets, 65 countries also possessed at least 500 unique names per gender in the WGND to enable sampling for validation and test sets.
For each of these 65 countries, we constructed a validation set and a test set independently by sampling 10,000 names (5,000 female and 5,000 male) from the WGND data. 
We drew these samples using separate random number generators for each set and country.
For both the validation and test sets, we performed the sampling with replacement weighted by the reported frequencies of each name-gender pair. 
This approach allows us to simulate a realistic population sample while ensuring a sufficient number of samples.

Overall, our benchmark construction process resulted in a total of 65 countries with complete benchmark sets across both Orbis and WGND datasets. The Orbis validation and test sets for each of these countries consist of 450 females and 450 males each. The WGND validation and test sets for each country consist of 5,000 females and 5,000 males each. Table \ref{table:s2} shows the number of executives in the training set constructed from the Orbis data for each country and gender.

\subsection{Naive Bayes Classifier}

To prepare the data for our classifier, we first converted each first name into a feature vector.
To ensure a consistent and comparable feature space across all countries, we tokenized every first name from the aggregated training corpus into character-level $n$-grams.
We set the maximum $n$-gram length to the length of the longest name within this training corpus.
We then transformed these tokenized names into high-dimensional vectors using a Term Frequency-Inverse Document Frequency (TF-IDF) weighting scheme, where each component represents the TF-IDF weight of a specific character $n$-gram.

Using these feature vectors, we trained a separate gender classifier for each of the 65 countries.
We chose the Naive Bayes algorithm, a probabilistic classification method based on Bayes' theorem, which operates under the assumption of conditional independence among features.
Specifically, we employed the Complement Naive Bayes (CNB) variant, available in the `scikit-learn' Python library \cite{rennie2003, scikit-learn}.
We selected CNB for its robustness in handling imbalanced datasets, as it estimates feature parameters by considering features from all classes except the one being evaluated.
Furthermore, to counteract any remaining gender imbalance in the training data, we applied class-balanced sample weights during the training process.

The CNB classifier utilizes a smoothing hyperparameter, $\alpha$, that addresses the zero-frequency problem and enhances the model's robustness against unseen data and class imbalance.
We tuned a country-specific smoothing hyperparameter, $\alpha_c$, of the CNB classifier for country $c$ using its validation set constructed from the Orbis data.
To this end, for each $\alpha_c \in \{0.001, 0.01, 0.1, 1, 10\}$, we evaluated the trained CNB classifier by calculating the Area Under the Receiver Operating Characteristic Curve (ROC AUC) on this validation set constructed from the Orbis data for each country.
We used the tuned $\alpha_c$ value that yielded the highest ROC AUC for each country (see Table S3).

To ensure the reliability of gender assignments, among the 65 countries with both Orbis and WGND validation sets available, we retained the 61 where the CNB classifier achieved an ROC AUC of at least 0.8 on both validation sets. 
This criterion led to the exclusion of four countries (Bangladesh, China, India, and South Korea) due to their lower ROC AUC values on the WGND validation set.

\subsection{Country-Specific Thresholding for Gender Assignment}

To ensure the high reliability of our gender assignments, we applied a country-specific confidence threshold, $\theta_c$, to probability estimates produced by our CNB classifier. 
We assign gender $g$ to a name associated with country $c$ if and only if an estimate of the posterior probability of that name's gender $g$ is $\theta_c$ or larger; otherwise, we do not assign it.

For each of the 61 countries, we determined these thresholds based on the classifiers' performance on both Orbis and WGND validation sets. 
We defined a target F1 score of 0.9 as a practical standard for high-confidence assignments. 
The F1 score was computed only on the subset of instances in the validation set meeting the confidence threshold (i.e., $p_g \geq \theta$), with the male class treated as positive.

To find $\theta_c$, we examined a range of candidate thresholds from 0.9 to 0.9999. 
For each country, we determined the smallest threshold that achieved an F1 score of at least 0.9 on the Orbis validation set ($\theta_c^\text{Orbis}$) and, separately, on the WGND validation set ($\theta_c^\text{WGND}$). 
If the F1 score did not reach 0.9 even at a threshold of 0.9999, we consistently set the respective threshold ($\theta_c^\text{Orbis}$ or $\theta_c^\text{WGND}$) to 0.9999 for that country.
The final confidence threshold, $\theta_c$, was set to the maximum of these two values ($\theta_c = \max(\theta_c^\text{Orbis}, \theta_c^\text{WGND})$). 
The resulting thresholds, $\{\theta_c\}_c$, along with the tuned smoothing hyperparameters, $\{\alpha_c\}_c$, for all 61 countries are listed in Table \ref{table:s3}.

For comparison, we also performed country-specific tuning of its confidence threshold, $\theta_c$, for a name-based gender classifier that uses cultural consensus theory (the CCT classifier~\cite{vanbuskirk2023}).
We followed the same procedure used for tuning $\theta_c$ of our CNB classifier for this purpose.
The resulting thresholds, $\{\theta_c\}_c$ for all 61 countries are listed in Table \ref{table:s4}.

\subsection{Benchmark Evaluation}

We evaluated our trained CNB classifier on both the Orbis and WGND test sets for each country by calculating the ROC AUC and F1 score. 
We also computed the gender assignment rate (GAR)---the fraction of gender-assigned test samples---separately for female and male samples ($\text{GAR}_{\text{f}}$ and $\text{GAR}_{\text{m}}$).
For comparison, we computed the same set of metrics for the CCT classifier.
For all these metrics, higher values indicate better performance.

Table~\ref{table:s5} presents the benchmark results for both classifiers across all 61 countries in the Orbis test sets. 
As expected, the CNB classifier generally outperforms or matches the CCT classifier, given that the CNB model is directly trained on Orbis data.
Across these 61 countries, the CNB classifier achieved (mean $\pm$ standard deviation) ROC AUCs of $0.979 \pm 0.034$, F1 scores of $0.973 \pm 0.026$, and gender assignment rates of $0.896 \pm 0.131$ for female and $0.907 \pm 0.117$ for male test samples.
In comparison, the CCT classifier achieved ROC AUCs of $0.960 \pm 0.048$, F1 scores of $0.970 \pm 0.029$, with corresponding gender assignment rates of $0.827 \pm 0.137$ for female and $0.864 \pm 0.112$ for male test samples.

Table~\ref{table:s6} presents the benchmark results for both classifiers across all 61 countries in the WGND test sets. 
In contrast to Orbis, the CNB classifier performs comparably to or worse than the CCT classifier on the WGND test sets. 
Across these 61 countries, the CNB classifier achieved (mean $\pm$ standard deviation) ROC AUCs of $0.938 \pm 0.049$, F1 scores of $0.943 \pm 0.036$, and gender assignment rates of $0.748 \pm 0.198$ for female and $0.729 \pm 0.211$ for male test samples.
In comparison, the CCT classifier achieved ROC AUCs of $0.987 \pm 0.020$, F1 scores of $0.992 \pm 0.013$, with corresponding gender assignment rates of $0.845 \pm 0.121$ for female and $0.839 \pm 0.104$ for male test samples.

The two classifiers exhibit complementary strengths tied to the nature of the underlying datasets. 
The CCT classifier demonstrates superior generalization performance on the WGND benchmark, which is largely derived from general population data. 
In contrast, our CNB classifier, trained on the Orbis database of international executives, often outperforms the CCT classifier on the Orbis test set.
This observed difference also illustrates the inherent difficulty of inferring gender from names across countries and emphasizes the importance of cross-dataset evaluation.

For the main analysis, we opted to use the CNB classifier. 
This decision was motivated by the specific nature of our target population: academic researchers. 
Researchers, much like the executives in the Orbis database, may constitute a highly mobile international population, a characteristic often termed ``academic mobility'' \cite{ackers2008}. 
We reasoned that the demographic and naming patterns of this group might be more accurately captured by the Orbis data than by the more static, country-specific general population data represented by WGND. 
Therefore, we judged the CNB classifier to be a suitable choice for this specific analytical context.

\section{Number of authors by gender and country}

Table \ref{table:s7} lists the number of gender-assigned authors for each of the 61 countries that passed our initial accuracy benchmarks. 
For the main analysis, however, we applied an additional filtering step to ensure that each country had a sufficiently large population for robust demographic analysis. 
Specifically, we retained only those countries with at least 1,000 authors for each gender.
This criterion led to the exclusion of three countries: Lesotho, Papua New Guinea, and Somalia. 
The final analysis, therefore, focuses on the remaining 58 countries.

\section{Analysis of Trends in Researcher Populations and Gender Balance}

\subsection{Researcher Populations}

This section details the methodology for comparing long-term trends in researcher population growth between the nine Arab countries and the ten reference countries.

For each country $c$, we began by computing $N_c(t)$, the total number of gender-assigned authors affiliated with country $c$ who published at least one paper in year $t \in [2000, 2023]$.
To analyze the long-term growth, we then transformed this population count into a log-linear trajectory, $y_c(t) = \log N_c(t)$, where $N_c(t) > 0$ for all $c$ and $t$ considered. 
For each country, we fit the following linear regression model to the log-transformed annual author counts:
\begin{align}
y_c(t) = \beta_{c, 0} + \beta_c t + \varepsilon_c(t),
\label{eq:s2}
\end{align}
where $\beta_{c, 0}$ is the intercept term, $\beta_c$ is the estimated slope representing the annual growth rate, and $\varepsilon_c(t)$ is the residual term.

We obtained the estimated slopes $\{\beta_c\}_c$ for each of the nine Arab countries and the ten reference countries. 
To account for potential autocorrelation in the time-series data, we computed standard errors (SEs) and $p$-values using a Newey-West estimator. 
Table \ref{table:s8} reports the estimated slope $\beta_c$, its corresponding SE and $p$-value, and the coefficient of determination $R^2$ for each country.

\subsection{Gender Balance}

This section details the methodology for examining long-term trends in gender balance, specifically focusing on the proportion of female authors for each country over the period 2000--2023.

For each country $c$, we computed $p_c(t)$, the proportion of female authors among all active authors in year $t$.
We then fit the following linear regression model to the annual time series $\{p_c(t)\}$ for each country to estimate the trend in female author representation:
\begin{align}
p_c(t) = \gamma_{c, 0} + \gamma_c t + \delta_c(t),
\label{eq:s3}
\end{align}
where $\gamma_{c, 0}$ is the intercept term, $\gamma_c$ is the estimated slope representing the average annual change in the proportion of female authors, and $\delta_c(t)$ is the residual term.

We obtained the estimated slopes $\{\gamma_c\}_c$ for each of the nine Arab countries and the ten reference countries.
To account for potential autocorrelation in the time-series data, we computed standard errors (SEs) and $p$-values using a Newey-West estimator.
Table \ref{table:s9} reports the estimated slope $\gamma_c$, its corresponding SE and $p$-value, and the coefficient of determination $R^2$ for each country.

\section{Inter-publication Interval Threshold by Country and Gender}

Table \ref{table:s10} shows the inter-publication interval thresholds, $\Delta_{\mathrm{IPI},c,g}$, by country $c$ and gender $g$ when the survival probability threshold is set to 1\%, 2\%, and 5\%.

\section{Analyses of Authors' Career Metrics}

We computed several career metrics for all authors active in 2023. 
In addition to our primary metric, cumulative productivity, we calculated the following for each author:
\begin{itemize}
    \item \textbf{Total productivity:} The total number of publications by the end of 2023.
    \item \textbf{Publishing-career length:} The duration in years from their first publication to the end of 2023.
    \item \textbf{Publication gap experience:} A binary indicator, true if the author has ever experienced an inter-publication interval longer than their corresponding $\Delta_{\text{IPI},c,g}$ threshold.
    \item \textbf{Returning author status:} A binary indicator, true if the author was inactive in 2022 but became active in 2023, and had published at least one paper before 2023.
\end{itemize}

We first calculated the Spearman's rank correlation between cumulative productivity, total productivity, and publishing-career length for each country and gender (see Table~\ref{table:s11}).
The Spearman's rank correlation between cumulative productivity and total productivity was, as expected, very high, given that both are based on publication counts. 
However, the correlation was not perfect, which reflects that cumulative productivity is sensitive to the continuity of publication activity. 
In contrast, the correlation between cumulative productivity and publishing-career length was only moderate. 
This weaker correlation suggests that cumulative productivity is not a simple proxy for the length of publishing-career but rather captures an author's recent publication momentum.

We further analyzed two metrics: the proportion of all active authors with a publication gap experience (see Table~\ref{table:s12}), and the proportion of returning authors within newly active authors (see Table~\ref{table:s13}).
First, we observed a contrast between country groups. 
For both metrics, the proportions were substantially higher in the reference countries than in the Arab countries. 
On average, the proportion of authors with a publication gap experience was substantially higher in the reference countries (15.6\% for female; 24.4\% for male) than in the Arab countries (6.6\% for female; 12.1\% for male).
Similarly, on average, the proportion of returning authors was substantially higher in reference countries (15.7\% for female; 25.4\% for male) than in Arab countries (5.7\% for female; 9.5\% for male).
Second, we found a consistent gender difference across both metrics. 
In 56 out of 58 countries, a larger proportion of male authors had a publication gap experience compared to their female counterparts. 
Likewise, in 57 out of 58 countries, the proportion of returning authors was higher for men than for women. 
This counter-intuitive result is considered a consequence of our gender-specific inter-publication interval thresholds, which are empirically derived and are longer for women in most countries (48 out of 58 countries; see Table \ref{table:s10}).

\section{Researcher Inflow and Gender Gap in Cumulative Productivity by Country in 2023}

Table \ref{table:s14} shows the researcher flow and the gender gap in cumulative productivity for each country in 2023.

\section{Compound Annual Growth Rate of the Number of Active Authors}

We define the compound annual growth rate (CAGR) of the number of active authors from year \(t_1\) to year \(t_2\) as \((n_2 / n_1)^{1 / (t_2 - t_1)} - 1\), where \(n_1\) and \(n_2\) denote the number of active authors in years \(t_1\) and \(t_2\), respectively.

\section{Analysis Across Research Domains}

\subsection{Methods}

We additionally conducted our analysis across the four high-level research domains defined by OpenAlex: ``Health Sciences'', ``Life Sciences'', ``Physical Sciences'', and ``Social Sciences''.
Following the methodology used for country of affiliation (Section 4.1), we assigned research domains to each author based on the frequency of research domains associated with their publications. 
Specifically, we identified the research domain(s) that appeared most frequently across an author's entire publication record. 
We retained only those authors whose research domain was uniquely determined (i.e., those with a single primary domain and no ties in frequency). 
For each combination of target year, country, and research domain, we constructed researcher population pyramids using the same criteria and methodology as those used to compute the researcher population pyramids shown in Fig.~2 in the main text.

\subsection{Results}

\begin{figure*}[p]
\centering
\includegraphics[width=1\textwidth]{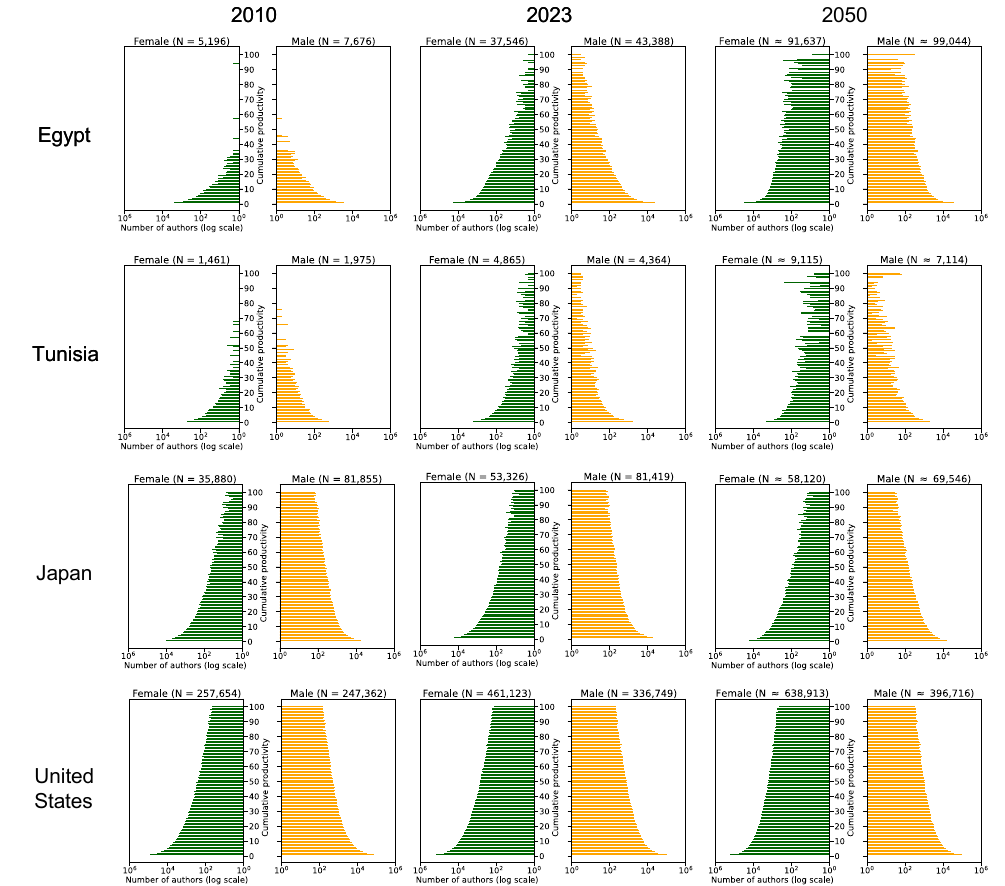}
\caption{Researcher population pyramids in 2010, 2023, and 2050 for Egypt, Tunisia, Japan, and the United States in Health Sciences. The number of active authors for each gender is displayed on a logarithmic horizontal axis. Total active author counts ($N$) are provided for each panel. The 2050 pyramids and their corresponding counts ($\approx$) are projections based on 2023 trends.}
\label{fig:s1}
\end{figure*}

\begin{figure*}[p]
\centering
\includegraphics[width=1\textwidth]{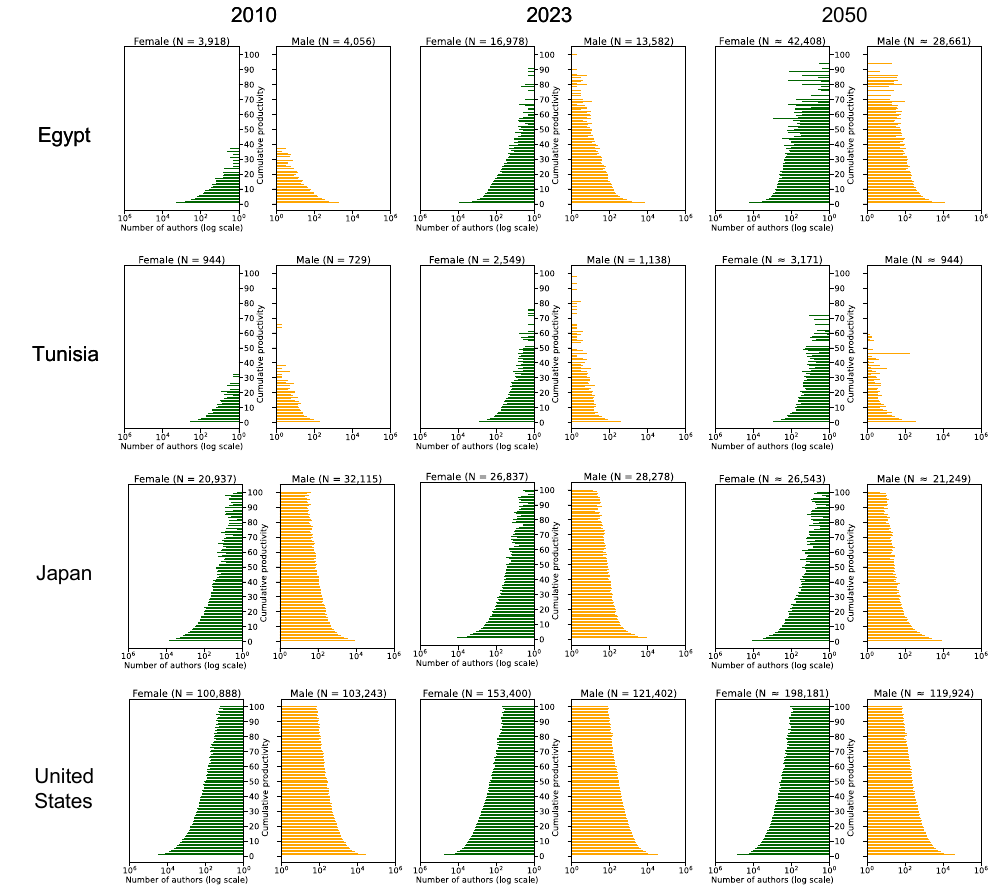}
\caption{Researcher population pyramids in 2010, 2023, and 2050 for Egypt, Tunisia, Japan, and the United States in Life Sciences. The number of active authors for each gender is displayed on a logarithmic horizontal axis. Total active author counts ($N$) are provided for each panel. The 2050 pyramids and their corresponding counts ($\approx$) are projections based on 2023 trends.}
\label{fig:s2}
\end{figure*}

\begin{figure*}[p]
\centering
\includegraphics[width=1\textwidth]{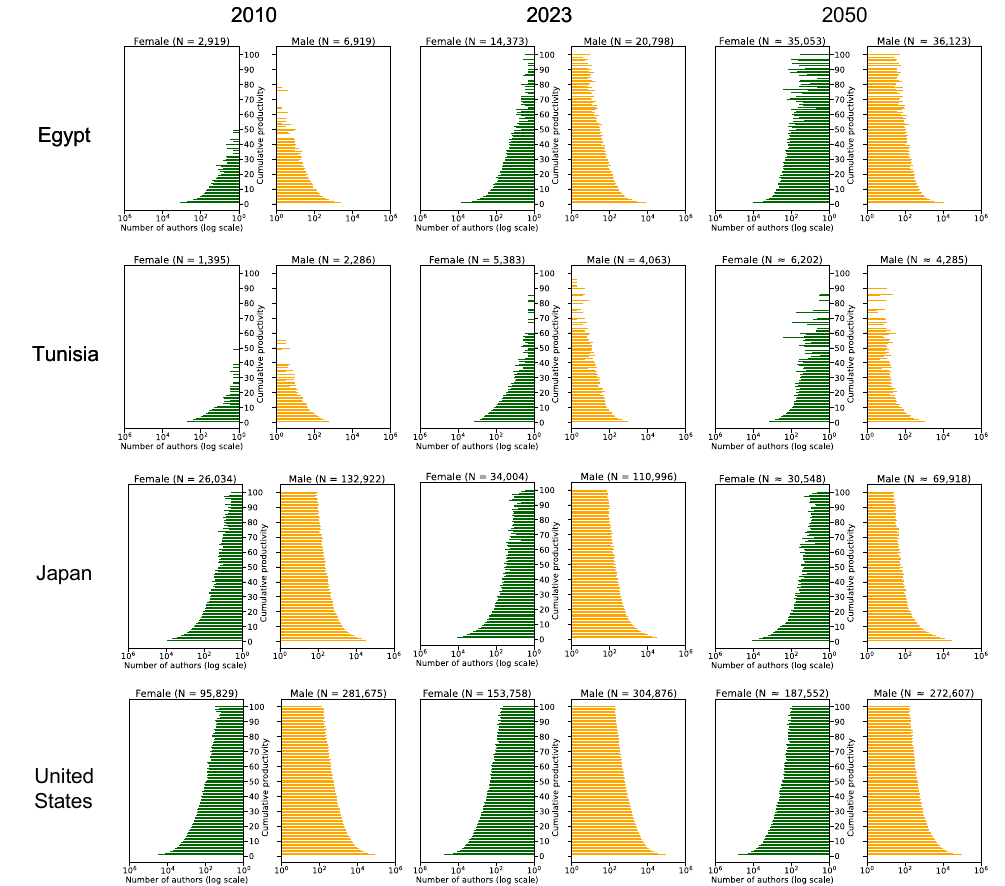}
\caption{Researcher population pyramids in 2010, 2023, and 2050 for Egypt, Tunisia, Japan, and the United States in Physical Sciences. The number of active authors for each gender is displayed on a logarithmic horizontal axis. Total active author counts ($N$) are provided for each panel. The 2050 pyramids and their corresponding counts ($\approx$) are projections based on 2023 trends.}
\label{fig:s3}
\end{figure*}

\begin{figure*}[p]
\centering
\includegraphics[width=1\textwidth]{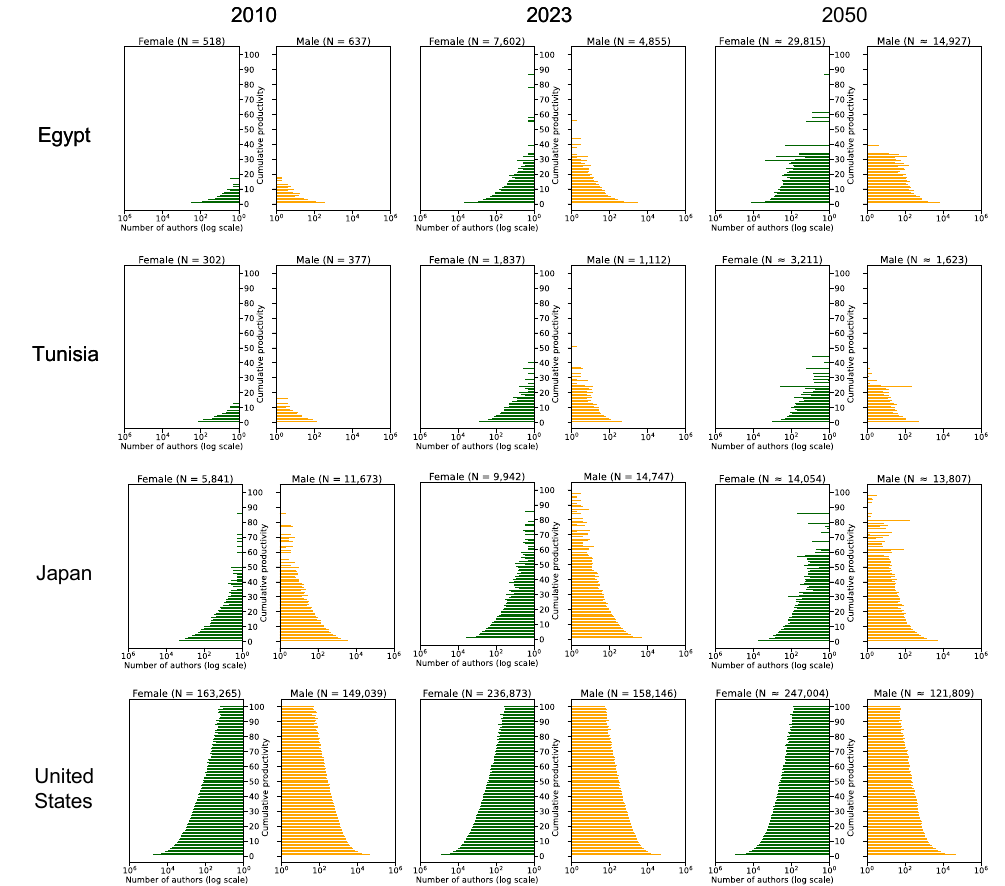}
\caption{Researcher population pyramids in 2010, 2023, and 2050 for Egypt, Tunisia, Japan, and the United States in Social Sciences. The number of active authors for each gender is displayed on a logarithmic horizontal axis. Total active author counts ($N$) are provided for each panel. The 2050 pyramids and their corresponding counts ($\approx$) are projections based on 2023 trends.}
\label{fig:s4}
\end{figure*}

Figures \ref{fig:s1}--\ref{fig:s4} present the researcher population pyramids for Egypt, Tunisia, Japan, and the United States across the four research domains for the years 2010, 2023, and 2050 (Health Sciences in Fig.~\ref{fig:s1}, Life Sciences in Fig.~\ref{fig:s2}, Physical Sciences in Fig.~\ref{fig:s3}, and Social Sciences in Fig.~\ref{fig:s4}).
Our analysis reveals two key findings. 
First, the vertical reach of the pyramids varies across domains, reflecting distinct disciplinary publication norms; for example, researchers in the Health Sciences exhibit higher maximum levels of cumulative productivity, while those in the Social Sciences exhibit lower overall publication volumes. 
Second, despite these differences in vertical scale, the qualitative characteristics of the three research systems remain largely consistent within each country across all four domains. 
Egypt and Tunisia consistently exhibit the ``Emerging'' pattern; the United States exhibits the ``Mature'' pattern; and Japan exhibits the ``Rigid'' pattern in every domain analyzed. 
These results indicate that while disciplinary norms influence the absolute height (cumulative productivity levels) of the pyramids, our framework captures the demographic trajectories of national research ecosystems.

\newpage

\begin{table}[h]
\centering
\caption{Member States of the Arab League \cite{arab_league}.}
\label{table:s1}
\begin{tabular}{l}
\hline
Country \\
\hline
Algeria \\
Bahrain \\
Comoros \\
Djibouti \\
Egypt \\
Iraq \\
Jordan \\
Kuwait \\
Lebanon \\
Libya \\
Mauritania \\
Morocco \\
Oman \\
Palestine, State of \\
Qatar \\
Saudi Arabia \\
Somalia \\
Sudan \\
Syrian Arab Republic \\
Tunisia \\
United Arab Emirates \\
Yemen \\
\hline
\end{tabular}
\end{table}

\newpage

\begin{longtable}{lrr}
\caption{Number of samples used in the training set for each country.} \\
\label{table:s2} \\
\hline
Country & Female & Male \\
\hline
\endfirsthead
\hline
Country & Female & Male \\
\hline
\endhead
Albania & 4,560 & 22,800\\
Algeria & 27,413 & 137,065\\
Australia & 45,789 & 104,429\\
Austria & 10,746 & 36,555\\
Bahrain & 35,066 & 45,000\\
Bangladesh & 5,515 & 22,171\\
Belarus & 4,385 & 12,505\\
Belgium & 8,741 & 32,846\\
Bosnia and Herzegovina & 2,741 & 8,511\\
Bulgaria & 609,226 & 822,719\\
Canada & 18,289 & 62,179\\
China & 90,973 & 222,418\\
Cyprus & 2,339 & 9,200\\
Czechia & 153,267 & 304,260\\
Denmark & 9,701 & 30,685\\
Egypt & 18,316 & 91,580\\
Estonia & 8,272 & 15,792\\
France & 61,820 & 170,251\\
Germany & 82,796 & 269,500\\
Ghana & 4,637 & 12,014\\
Iceland & 22,307 & 46,856\\
India & 94,220 & 305,744\\
Iran, Islamic Republic of & 7,050 & 35,250\\
Iraq & 273,599 & 1,041,777\\
Ireland & 70,759 & 149,920\\
Israel & 4,357 & 21,785\\
Italy & 2,508,071 & 5,348,406\\
Jamaica & 4,754 & 7,588\\
Japan & 9,200 & 46,000\\
Jordan & 71,967 & 359,835\\
Kenya & 16,919 & 59,825\\
Korea, Republic of & 4,781 & 23,905\\
Kuwait & 14,005 & 49,845\\
Lebanon & 55,521 & 130,120\\
Lesotho & 8,961 & 14,983\\
Lithuania & 11,419 & 26,476\\
Moldova, Republic of & 20,717 & 39,152\\
Montenegro & 5,990 & 13,186\\
Morocco & 3,075 & 15,375\\
Netherlands & 21,159 & 77,841\\
New Zealand & 314,854 & 517,150\\
Nigeria & 26,410 & 55,444\\
Norway & 425,903 & 694,519\\
Oman & 33,653 & 140,042\\
Papua New Guinea & 35,812 & 114,524\\
Philippines & 61,500 & 40,699\\
Poland & 55,843 & 133,137\\
Portugal & 18,029 & 44,429\\
Romania & 611,956 & 948,609\\
Russian Federation & 6,808,203 & 10,064,454\\
Saudi Arabia & 5,111 & 25,555\\
Serbia & 174,091 & 330,888\\
Somalia & 273,599 & 1,041,777\\
South Africa & 3,060,324 & 3,829,095\\
Spain & 23,016 & 60,053\\
Sri Lanka & 4,841 & 13,854\\
Sweden & 24,125 & 62,170\\
Switzerland & 376,575 & 746,801\\
Syrian Arab Republic & 273,599 & 1,041,777\\
Tunisia & 273,599 & 1,041,777\\
Turkey & 17,420 & 87,100\\
United Arab Emirates & 9,472 & 47,360\\
United Kingdom & 4,735,568 & 7,910,437\\
United States & 70,070 & 259,774\\
Zimbabwe & 8,837 & 11,148\\
\hline
\end{longtable}

\newpage

\begin{longtable}{lcc}
\caption{Country-specific tuned hyperparameters of our CNB classifier.} \\
\label{table:s3} \\
\hline
Country & $\alpha_c$ & $\theta_c$ \\
\hline\endfirsthead
\hline
Country & $\alpha_c$ & $\theta_c$ \\
\hline\endhead
Albania & $0.100$ & $0.9000$\\
Algeria & $1.000$ & $0.9000$\\
Australia & $0.010$ & $0.9000$\\
Austria & $0.001$ & $0.9700$\\
Bahrain & $0.010$ & $0.9930$\\
Belarus & $0.001$ & $0.9992$\\
Belgium & $0.100$ & $0.9000$\\
Bosnia and Herzegovina & $0.100$ & $0.9000$\\
Bulgaria & $0.001$ & $0.9000$\\
Canada & $0.100$ & $0.9000$\\
Cyprus & $0.010$ & $0.9900$\\
Czechia & $10.000$ & $0.9000$\\
Denmark & $0.100$ & $0.9000$\\
Egypt & $0.100$ & $0.9000$\\
Estonia & $0.100$ & $0.9000$\\
France & $0.100$ & $0.9000$\\
Germany & $1.000$ & $0.9000$\\
Ghana & $0.100$ & $0.9000$\\
Iceland & $0.010$ & $0.9000$\\
Iran, Islamic Republic of & $0.100$ & $0.9400$\\
Iraq & $0.100$ & $0.9998$\\
Ireland & $0.100$ & $0.9000$\\
Israel & $0.100$ & $0.9000$\\
Italy & $0.001$ & $0.9000$\\
Jamaica & $0.100$ & $0.9000$\\
Japan & $0.100$ & $0.9000$\\
Jordan & $0.100$ & $0.9300$\\
Kenya & $0.100$ & $0.9400$\\
Kuwait & $0.010$ & $0.9800$\\
Lebanon & $0.010$ & $0.9800$\\
Lesotho & $0.100$ & $0.9000$\\
Lithuania & $0.001$ & $0.9000$\\
Moldova, Republic of & $0.001$ & $0.9900$\\
Montenegro & $0.100$ & $0.9000$\\
Morocco & $1.000$ & $0.9000$\\
Netherlands & $0.100$ & $0.9000$\\
New Zealand & $0.100$ & $0.9000$\\
Nigeria & $0.100$ & $0.9000$\\
Norway & $0.100$ & $0.9000$\\
Oman & $0.010$ & $0.9960$\\
Papua New Guinea & $0.100$ & $0.9000$\\
Philippines & $1.000$ & $0.9000$\\
Poland & $0.001$ & $0.9940$\\
Portugal & $0.100$ & $0.9000$\\
Romania & $10.000$ & $0.9000$\\
Russian Federation & $0.001$ & $0.9400$\\
Saudi Arabia & $0.100$ & $0.9500$\\
Serbia & $0.001$ & $0.9600$\\
Somalia & $0.010$ & $0.9999$\\
South Africa & $0.010$ & $0.9000$\\
Spain & $0.100$ & $0.9000$\\
Sri Lanka & $1.000$ & $0.9000$\\
Sweden & $0.100$ & $0.9000$\\
Switzerland & $0.010$ & $0.9000$\\
Syrian Arab Republic & $0.100$ & $0.9300$\\
Tunisia & $0.100$ & $0.9000$\\
T\"{u}rkiye & $0.100$ & $0.9000$\\
United Arab Emirates & $0.001$ & $0.9970$\\
United Kingdom & $1.000$ & $0.9000$\\
United States & $0.100$ & $0.9000$\\
Zimbabwe & $0.010$ & $0.9000$\\
\hline
\end{longtable}

\newpage

\begin{longtable}{lcc}
\caption{Country-specific tuned hyperparameters of the CCT classifier.} \\
\label{table:s4} \\
\hline
Country & $\theta_c$ \\
\hline\endfirsthead
\hline
Country & $\theta_c$ \\
\hline\endhead
Albania & $0.9000$\\
Algeria & $0.9000$\\
Australia & $0.9000$\\
Austria & $0.9000$\\
Bahrain & $0.9000$\\
Belarus & $0.9000$\\
Belgium & $0.9000$\\
Bosnia and Herzegovina & $0.9000$\\
Bulgaria & $0.9000$\\
Canada & $0.9000$\\
Cyprus & $0.9000$\\
Czechia & $0.9000$\\
Denmark & $0.9000$\\
Egypt & $0.9000$\\
Estonia & $0.9000$\\
France & $0.9000$\\
Germany & $0.9000$\\
Ghana & $0.9000$\\
Iceland & $0.9000$\\
Iran, Islamic Republic of & $0.9000$\\
Iraq & $0.9000$\\
Ireland & $0.9000$\\
Israel & $0.9000$\\
Italy & $0.9000$\\
Jamaica & $0.9000$\\
Japan & $0.9000$\\
Jordan & $0.9000$\\
Kenya & $0.9000$\\
Kuwait & $0.9000$\\
Lebanon & $0.9000$\\
Lesotho & $0.9999$\\
Lithuania & $0.9000$\\
Moldova, Republic of & $0.9000$\\
Montenegro & $0.9000$\\
Morocco & $0.9000$\\
Netherlands & $0.9000$\\
New Zealand & $0.9000$\\
Nigeria & $0.9000$\\
Norway & $0.9000$\\
Oman & $0.9000$\\
Papua New Guinea & $0.9000$\\
Philippines & $0.9000$\\
Poland & $0.9000$\\
Portugal & $0.9000$\\
Romania & $0.9000$\\
Russian Federation & $0.9000$\\
Saudi Arabia & $0.9300$\\
Serbia & $0.9000$\\
Somalia & $0.9000$\\
South Africa & $0.9000$\\
Spain & $0.9000$\\
Sri Lanka & $0.9000$\\
Sweden & $0.9000$\\
Switzerland & $0.9000$\\
Syrian Arab Republic & $0.9000$\\
Tunisia & $0.9000$\\
T\"{u}rkiye & $0.9000$\\
United Arab Emirates & $0.9000$\\
United Kingdom & $0.9000$\\
United States & $0.9000$\\
Zimbabwe & $0.9000$\\
\hline
\end{longtable}

\newpage

\begin{longtable}{lcccc|cccc}
\caption{
Benchmark results on the Orbis test sets for the CNB and CCT classifiers. 
Columns AUC and F1 represent the ROC AUC and F1 score, respectively. 
$\text{GAR}_{\text{f}}$ and $\text{GAR}_{\text{m}}$ represent the gender assignment rates for the female and male test samples, respectively.} \\
\label{table:s5} \\
\hline
 & \multicolumn{4}{c|}{CNB} & \multicolumn{4}{c}{CCT} \\
Country & AUC & F1 & $\text{GAR}_{\text{f}}$ & $\text{GAR}_{\text{m}}$ & AUC & F1 & $\text{GAR}_{\text{f}}$ & $\text{GAR}_{\text{m}}$ \\
\hline\endfirsthead
\hline
 & \multicolumn{4}{c|}{CNB} & \multicolumn{4}{c}{CCT} \\
Country & AUC & F1 & $\text{GAR}_{\text{f}}$ & $\text{GAR}_{\text{m}}$ & AUC & F1 & $\text{GAR}_{\text{f}}$ & $\text{GAR}_{\text{m}}$ \\
\hline\endhead
Albania & 0.984 & 0.963 & 0.902 & 0.911 & 0.950 & 0.963 & 0.853 & 0.869 \\
Algeria & 0.965 & 0.951 & 0.880 & 0.931 & 0.945 & 0.953 & 0.847 & 0.911 \\
Australia & 0.995 & 0.978 & 0.964 & 0.947 & 0.992 & 0.994 & 0.891 & 0.931 \\
Austria & 0.996 & 0.984 & 0.962 & 0.978 & 0.995 & 0.996 & 0.927 & 0.967 \\
Bahrain & 0.998 & 0.998 & 0.960 & 0.956 & 0.969 & 0.969 & 0.824 & 0.891 \\
Belarus & 0.999 & 0.999 & 0.951 & 0.980 & 0.997 & 0.998 & 0.989 & 0.967 \\
Belgium & 0.982 & 0.971 & 0.887 & 0.873 & 0.979 & 0.985 & 0.896 & 0.864 \\
Bosnia and Herzegovina & 0.986 & 0.974 & 0.889 & 0.927 & 0.979 & 0.978 & 0.896 & 0.920 \\
Bulgaria & 0.998 & 0.993 & 0.996 & 0.984 & 0.987 & 0.991 & 0.933 & 0.798 \\
Canada & 0.987 & 0.973 & 0.893 & 0.918 & 0.984 & 0.986 & 0.851 & 0.918 \\
Cyprus & 0.976 & 0.974 & 0.749 & 0.889 & 0.971 & 0.970 & 0.847 & 0.931 \\
Czechia & 0.999 & 0.995 & 0.967 & 0.951 & 0.999 & 0.999 & 0.953 & 0.909 \\
Denmark & 0.989 & 0.977 & 0.884 & 0.931 & 0.987 & 0.987 & 0.864 & 0.893 \\
Egypt & 0.988 & 0.968 & 0.924 & 0.953 & 0.945 & 0.944 & 0.733 & 0.849 \\
Estonia & 0.998 & 0.994 & 0.956 & 0.969 & 0.991 & 0.992 & 0.847 & 0.851 \\
France & 0.995 & 0.986 & 0.960 & 0.916 & 0.991 & 0.995 & 0.902 & 0.858 \\
Germany & 0.995 & 0.991 & 0.944 & 0.933 & 0.995 & 0.995 & 0.931 & 0.940 \\
Ghana & 0.994 & 0.982 & 0.938 & 0.920 & 0.975 & 0.986 & 0.876 & 0.896 \\
Iceland & 0.999 & 0.997 & 0.984 & 0.987 & 0.988 & 0.990 & 0.902 & 0.918 \\
Iran, Islamic Republic of & 0.981 & 0.974 & 0.860 & 0.844 & 0.933 & 0.941 & 0.769 & 0.856 \\
Iraq & 0.841 & 0.938 & 0.347 & 0.364 & 0.906 & 0.927 & 0.618 & 0.780 \\
Ireland & 0.999 & 0.997 & 0.958 & 0.971 & 0.998 & 0.998 & 0.918 & 0.969 \\
Israel & 0.965 & 0.953 & 0.822 & 0.889 & 0.941 & 0.952 & 0.811 & 0.820 \\
Italy & 0.999 & 0.999 & 0.998 & 0.999 & 0.999 & 0.999 & 0.991 & 0.860 \\
Jamaica & 0.978 & 0.969 & 0.838 & 0.833 & 0.964 & 0.972 & 0.762 & 0.816 \\
Japan & 0.981 & 0.956 & 0.904 & 0.964 & 0.967 & 0.970 & 0.733 & 0.882 \\
Jordan & 0.995 & 0.986 & 0.971 & 0.958 & 0.951 & 0.954 & 0.736 & 0.867 \\
Kenya & 0.914 & 0.895 & 0.918 & 0.922 & 0.907 & 0.896 & 0.938 & 0.929 \\
Kuwait & 0.985 & 0.968 & 0.898 & 0.927 & 0.928 & 0.942 & 0.631 & 0.876 \\
Lebanon & 0.997 & 0.992 & 0.942 & 0.976 & 0.926 & 0.931 & 0.802 & 0.929 \\
Lesotho & 0.975 & 0.965 & 0.887 & 0.847 & 0.687 & 0.858 & 0.147 & 0.213 \\
Lithuania & 0.994 & 0.991 & 0.998 & 0.993 & 0.994 & 0.994 & 0.920 & 0.971 \\
Moldova, Republic of & 0.993 & 0.986 & 0.987 & 0.960 & 0.985 & 0.986 & 0.969 & 0.913 \\
Montenegro & 0.983 & 0.955 & 0.916 & 0.920 & 0.955 & 0.959 & 0.864 & 0.889 \\
Morocco & 0.989 & 0.985 & 0.898 & 0.893 & 0.973 & 0.976 & 0.911 & 0.913 \\
Netherlands & 0.981 & 0.973 & 0.864 & 0.931 & 0.980 & 0.984 & 0.864 & 0.896 \\
New Zealand & 0.998 & 0.992 & 0.962 & 0.942 & 0.989 & 0.991 & 0.898 & 0.893 \\
Nigeria & 0.954 & 0.923 & 0.862 & 0.869 & 0.933 & 0.952 & 0.604 & 0.742 \\
Norway & 0.991 & 0.980 & 0.980 & 0.987 & 0.984 & 0.986 & 0.893 & 0.887 \\
Oman & 0.999 & 0.997 & 0.958 & 0.980 & 0.908 & 0.918 & 0.742 & 0.851 \\
Papua New Guinea & 0.983 & 0.969 & 0.887 & 0.907 & 0.946 & 0.970 & 0.764 & 0.784 \\
Philippines & 0.981 & 0.978 & 0.873 & 0.849 & 0.958 & 0.968 & 0.784 & 0.822 \\
Poland & 0.992 & 0.990 & 0.982 & 0.973 & 0.991 & 0.993 & 0.973 & 0.929 \\
Portugal & 0.996 & 0.992 & 0.933 & 0.913 & 0.992 & 0.994 & 0.924 & 0.924 \\
Romania & 0.980 & 0.967 & 0.922 & 0.920 & 0.962 & 0.973 & 0.851 & 0.787 \\
Russian Federation & 0.999 & 0.996 & 0.993 & 0.998 & 0.998 & 0.999 & 0.964 & 0.929 \\
Saudi Arabia & 0.981 & 0.975 & 0.858 & 0.933 & 0.878 & 0.901 & 0.684 & 0.849 \\
Serbia & 0.997 & 0.994 & 0.967 & 0.987 & 0.984 & 0.979 & 0.949 & 0.938 \\
Somalia & 0.824 & 0.883 & 0.280 & 0.480 & 0.964 & 0.971 & 0.749 & 0.867 \\
South Africa & 0.979 & 0.962 & 0.884 & 0.858 & 0.919 & 0.966 & 0.624 & 0.671 \\
Spain & 0.990 & 0.978 & 0.944 & 0.940 & 0.986 & 0.984 & 0.920 & 0.924 \\
Sri Lanka & 0.884 & 0.927 & 0.500 & 0.473 & 0.858 & 0.930 & 0.562 & 0.582 \\
Sweden & 0.996 & 0.988 & 0.922 & 0.949 & 0.988 & 0.992 & 0.933 & 0.909 \\
Switzerland & 0.999 & 0.992 & 0.989 & 0.991 & 0.999 & 0.999 & 0.962 & 0.933 \\
Syrian Arab Republic & 0.917 & 0.888 & 0.822 & 0.840 & 0.943 & 0.954 & 0.653 & 0.853 \\
Tunisia & 0.965 & 0.953 & 0.898 & 0.844 & 0.964 & 0.966 & 0.807 & 0.871 \\
T\"{u}rkiye & 0.990 & 0.975 & 0.924 & 0.936 & 0.972 & 0.977 & 0.811 & 0.898 \\
United Arab Emirates & 0.995 & 0.991 & 0.933 & 0.969 & 0.942 & 0.950 & 0.742 & 0.902 \\
United Kingdom & 0.998 & 0.991 & 0.967 & 0.964 & 0.996 & 0.998 & 0.876 & 0.918 \\
United States & 0.987 & 0.968 & 0.942 & 0.944 & 0.978 & 0.984 & 0.849 & 0.933 \\
Zimbabwe & 0.970 & 0.952 & 0.891 & 0.862 & 0.941 & 0.972 & 0.771 & 0.658 \\
\hline
\end{longtable}

\newpage

\begin{longtable}{lcccc|cccc}
\caption{Benchmark results on the WGND test sets for the CNB and CCT classifiers. 
Columns AUC and F1 represent the ROC AUC and F1 score, respectively. 
$\text{GAR}_{\text{f}}$ and $\text{GAR}_{\text{m}}$ represent the gender assignment rates for the female and male test samples, respectively.
} \\
\label{table:s6} \\
\hline
 & \multicolumn{4}{c|}{CNB} & \multicolumn{4}{c}{CCT} \\
Country & AUC & F1 & $\text{GAR}_{\text{f}}$ & $\text{GAR}_{\text{m}}$ & AUC & F1 & $\text{GAR}_{\text{f}}$ & $\text{GAR}_{\text{m}}$ \\
\hline\endfirsthead
\hline
 & \multicolumn{4}{c|}{CNB} & \multicolumn{4}{c}{CCT} \\
Country & AUC & F1 & $\text{GAR}_{\text{f}}$ & $\text{GAR}_{\text{m}}$ & AUC & F1 & $\text{GAR}_{\text{f}}$ & $\text{GAR}_{\text{m}}$ \\
\hline\endhead
Albania & 0.934 & 0.918 & 0.781 & 0.842 & 0.997 & 0.998 & 0.916 & 0.855 \\
Algeria & 0.871 & 0.938 & 0.468 & 0.359 & 0.988 & 0.996 & 0.796 & 0.832 \\
Australia & 0.993 & 0.979 & 0.956 & 0.959 & 0.996 & 0.996 & 0.893 & 0.928 \\
Austria & 0.915 & 0.899 & 0.797 & 0.812 & 0.992 & 0.993 & 0.908 & 0.881 \\
Bahrain & 0.908 & 0.913 & 0.747 & 0.555 & 0.990 & 0.996 & 0.809 & 0.843 \\
Belarus & 0.859 & 0.907 & 0.439 & 0.414 & 0.902 & 0.936 & 0.708 & 0.648 \\
Belgium & 0.991 & 0.984 & 0.894 & 0.885 & 0.998 & 0.998 & 0.920 & 0.893 \\
Bosnia and Herzegovina & 0.962 & 0.954 & 0.762 & 0.781 & 0.986 & 0.991 & 0.904 & 0.882 \\
Bulgaria & 0.999 & 0.999 & 0.999 & 0.999 & 0.999 & 0.999 & 0.999 & 0.820 \\
Canada & 0.993 & 0.981 & 0.948 & 0.939 & 0.998 & 0.998 & 0.885 & 0.934 \\
Cyprus & 0.824 & 0.906 & 0.320 & 0.267 & 0.992 & 0.991 & 0.917 & 0.856 \\
Czechia & 0.959 & 0.965 & 0.768 & 0.670 & 0.970 & 0.975 & 0.847 & 0.783 \\
Denmark & 0.999 & 0.999 & 0.997 & 0.956 & 0.999 & 0.999 & 0.915 & 0.956 \\
Egypt & 0.895 & 0.904 & 0.649 & 0.601 & 0.988 & 0.996 & 0.814 & 0.834 \\
Estonia & 0.982 & 0.970 & 0.872 & 0.917 & 0.996 & 0.999 & 0.823 & 0.874 \\
France & 0.999 & 0.997 & 0.987 & 0.961 & 0.999 & 0.999 & 0.936 & 0.855 \\
Germany & 0.933 & 0.937 & 0.772 & 0.667 & 0.993 & 0.993 & 0.894 & 0.883 \\
Ghana & 0.906 & 0.923 & 0.673 & 0.590 & 0.992 & 0.989 & 0.914 & 0.854 \\
Iceland & 0.993 & 0.975 & 0.922 & 0.966 & 0.983 & 0.993 & 0.737 & 0.815 \\
Iran, Islamic Republic of & 0.857 & 0.892 & 0.526 & 0.516 & 0.996 & 0.998 & 0.856 & 0.893 \\
Iraq & 0.944 & 0.977 & 0.508 & 0.342 & 0.988 & 0.995 & 0.808 & 0.834 \\
Ireland & 0.999 & 0.993 & 0.983 & 0.976 & 0.999 & 0.999 & 0.930 & 0.958 \\
Israel & 0.937 & 0.933 & 0.734 & 0.759 & 0.988 & 0.991 & 0.852 & 0.800 \\
Italy & 0.995 & 0.991 & 0.985 & 0.988 & 0.999 & 0.999 & 0.946 & 0.910 \\
Jamaica & 0.909 & 0.929 & 0.608 & 0.576 & 0.992 & 0.990 & 0.907 & 0.852 \\
Japan & 0.928 & 0.908 & 0.759 & 0.811 & 0.945 & 0.970 & 0.625 & 0.726 \\
Jordan & 0.901 & 0.896 & 0.737 & 0.647 & 0.990 & 0.997 & 0.809 & 0.831 \\
Kenya & 0.886 & 0.906 & 0.516 & 0.633 & 0.992 & 0.991 & 0.910 & 0.847 \\
Kuwait & 0.888 & 0.901 & 0.655 & 0.556 & 0.990 & 0.996 & 0.808 & 0.836 \\
Lebanon & 0.897 & 0.906 & 0.693 & 0.597 & 0.990 & 0.996 & 0.807 & 0.838 \\
Lesotho & 0.846 & 0.908 & 0.375 & 0.474 & 0.992 & 0.994 & 0.089 & 0.259 \\
Lithuania & 0.953 & 0.927 & 0.909 & 0.942 & 0.982 & 0.991 & 0.837 & 0.829 \\
Moldova, Republic of & 0.871 & 0.908 & 0.559 & 0.524 & 0.896 & 0.933 & 0.689 & 0.638 \\
Montenegro & 0.888 & 0.932 & 0.439 & 0.526 & 0.993 & 0.994 & 0.912 & 0.888 \\
Morocco & 0.863 & 0.948 & 0.353 & 0.284 & 0.989 & 0.997 & 0.795 & 0.829 \\
Netherlands & 0.971 & 0.968 & 0.826 & 0.787 & 0.995 & 0.994 & 0.837 & 0.839 \\
New Zealand & 0.999 & 0.988 & 0.971 & 0.987 & 0.999 & 0.999 & 0.899 & 0.950 \\
Nigeria & 0.906 & 0.922 & 0.629 & 0.687 & 0.991 & 0.991 & 0.910 & 0.855 \\
Norway & 0.999 & 0.993 & 0.954 & 0.971 & 0.999 & 0.999 & 0.879 & 0.929 \\
Oman & 0.876 & 0.902 & 0.593 & 0.495 & 0.988 & 0.996 & 0.812 & 0.825 \\
Papua New Guinea & 0.939 & 0.918 & 0.839 & 0.824 & 0.992 & 0.990 & 0.911 & 0.854 \\
Philippines & 0.999 & 0.999 & 0.912 & 0.979 & 0.996 & 0.990 & 0.919 & 0.999 \\
Poland & 0.910 & 0.906 & 0.688 & 0.738 & 0.966 & 0.989 & 0.811 & 0.660 \\
Portugal & 0.974 & 0.961 & 0.864 & 0.862 & 0.999 & 0.999 & 0.956 & 0.961 \\
Romania & 0.931 & 0.939 & 0.591 & 0.566 & 0.987 & 0.987 & 0.906 & 0.830 \\
Russian Federation & 0.934 & 0.895 & 0.864 & 0.872 & 0.923 & 0.958 & 0.722 & 0.683 \\
Saudi Arabia & 0.888 & 0.903 & 0.645 & 0.488 & 0.987 & 0.996 & 0.768 & 0.799 \\
Serbia & 0.923 & 0.905 & 0.727 & 0.791 & 0.989 & 0.989 & 0.914 & 0.867 \\
Somalia & 0.945 & 0.975 & 0.551 & 0.400 & 0.989 & 0.996 & 0.803 & 0.837 \\
South Africa & 0.978 & 0.954 & 0.913 & 0.897 & 0.991 & 0.991 & 0.896 & 0.859 \\
Spain & 0.997 & 0.992 & 0.950 & 0.946 & 0.992 & 0.999 & 0.844 & 0.857 \\
Sri Lanka & 0.909 & 0.966 & 0.201 & 0.406 & 0.989 & 0.995 & 0.808 & 0.744 \\
Sweden & 0.998 & 0.993 & 0.946 & 0.953 & 0.999 & 0.999 & 0.920 & 0.918 \\
Switzerland & 0.997 & 0.988 & 0.955 & 0.964 & 0.999 & 0.999 & 0.943 & 0.910 \\
Syrian Arab Republic & 0.943 & 0.918 & 0.874 & 0.713 & 0.989 & 0.996 & 0.803 & 0.835 \\
Tunisia & 0.943 & 0.913 & 0.889 & 0.747 & 0.988 & 0.996 & 0.803 & 0.834 \\
T\"{u}rkiye & 0.983 & 0.970 & 0.887 & 0.889 & 0.994 & 0.994 & 0.835 & 0.853 \\
United Arab Emirates & 0.878 & 0.906 & 0.603 & 0.544 & 0.990 & 0.997 & 0.816 & 0.832 \\
United Kingdom & 0.995 & 0.984 & 0.951 & 0.930 & 0.998 & 0.998 & 0.889 & 0.882 \\
United States & 0.989 & 0.972 & 0.952 & 0.932 & 0.995 & 0.993 & 0.887 & 0.928 \\
Zimbabwe & 0.923 & 0.906 & 0.789 & 0.780 & 0.992 & 0.990 & 0.912 & 0.852 \\
\hline
\end{longtable}

\newpage

\begin{longtable}{l rrr cc} 
\caption{Number of authors by gender, proportion of female authors, and gender assignment rate for each country. 
$N$: total number of authors with an assigned country. 
$n_{\text{female}}$: number of female authors. 
$n_{\text{male}}$: number of male authors. 
$p_{\text{female}}$: proportion of female authors among gender-assigned authors ($n_{\text{female}} / (n_{\text{female}} + n_{\text{male}})$). 
GAR: gender assignment rate (($n_{\text{female}} + n_{\text{male}}) / N$).
} \\
\label{table:s7} \\
\hline
Country & \multicolumn{1}{c}{$N$} & \multicolumn{1}{c}{$n_{\text{female}}$} & \multicolumn{1}{c}{$n_{\text{male}}$} & $p_{\text{female}}$ (\%) & GAR (\%) \\
\hline\endfirsthead
\hline
Country & \multicolumn{1}{c}{$N$} & \multicolumn{1}{c}{$n_{\text{female}}$} & \multicolumn{1}{c}{$n_{\text{male}}$} & $p_{\text{female}}$ (\%) & GAR (\%) \\
\hline\endhead
Albania & 8,101 & 3,636 & 2,484 & 59.4\% & 75.5\%\\
Algeria & 82,032 & 24,186 & 28,860 & 45.6\% & 64.7\%\\
Australia & 449,116 & 157,027 & 193,939 & 44.7\% & 78.1\%\\
Austria & 140,857 & 43,527 & 64,650 & 40.2\% & 76.8\%\\
Bahrain & 6,383 & 2,377 & 2,163 & 52.4\% & 71.1\%\\
Belarus & 32,530 & 3,665 & 4,458 & 45.1\% & 25.0\%\\
Belgium & 193,359 & 56,615 & 75,125 & 43.0\% & 68.1\%\\
Bosnia and Herzegovina & 15,878 & 6,959 & 5,584 & 55.5\% & 79.0\%\\
Bulgaria & 46,690 & 17,182 & 20,643 & 45.4\% & 81.0\%\\
Canada & 709,352 & 232,752 & 266,578 & 46.6\% & 70.4\%\\
Cyprus & 13,567 & 3,162 & 4,876 & 39.3\% & 59.2\%\\
Czechia & 138,070 & 40,978 & 55,632 & 42.4\% & 70.0\%\\
Denmark & 137,867 & 47,737 & 54,154 & 46.9\% & 73.9\%\\
Egypt & 273,642 & 97,512 & 110,256 & 46.9\% & 75.9\%\\
Estonia & 18,136 & 6,826 & 6,224 & 52.3\% & 72.0\%\\
France & 975,619 & 280,487 & 429,656 & 39.5\% & 72.8\%\\
Germany & 1,362,026 & 342,417 & 551,829 & 38.3\% & 65.7\%\\
Ghana & 35,647 & 7,631 & 19,671 & 28.0\% & 76.6\%\\
Iceland & 7,445 & 2,977 & 3,382 & 46.8\% & 85.4\%\\
Iran, Islamic Republic of & 580,434 & 208,756 & 246,589 & 45.8\% & 78.4\%\\
Iraq & 115,842 & 25,416 & 27,660 & 47.9\% & 45.8\%\\
Ireland & 74,655 & 26,894 & 28,408 & 48.6\% & 74.1\%\\
Israel & 165,123 & 51,135 & 69,100 & 42.5\% & 72.8\%\\
Italy & 671,119 & 239,535 & 294,198 & 44.9\% & 79.5\%\\
Jamaica & 3,938 & 1,194 & 995 & 54.5\% & 55.6\%\\
Japan & 1,426,766 & 268,393 & 849,383 & 24.0\% & 78.3\%\\
Jordan & 41,137 & 13,926 & 19,738 & 41.4\% & 81.8\%\\
Kenya & 51,264 & 12,196 & 22,554 & 35.1\% & 67.8\%\\
Kuwait & 17,140 & 4,271 & 6,733 & 38.8\% & 64.2\%\\
Lebanon & 21,983 & 9,538 & 7,541 & 55.8\% & 77.7\%\\
Lesotho & 826 & 219 & 262 & 45.5\% & 58.2\%\\
Lithuania & 32,502 & 16,170 & 12,388 & 56.6\% & 87.9\%\\
Moldova, Republic of & 9,729 & 3,510 & 3,164 & 52.6\% & 68.6\%\\
Montenegro & 2,679 & 1,148 & 942 & 54.9\% & 78.0\%\\
Morocco & 77,995 & 17,994 & 26,155 & 40.8\% & 56.6\%\\
Netherlands & 394,621 & 107,770 & 160,882 & 40.1\% & 68.1\%\\
New Zealand & 75,098 & 25,651 & 31,850 & 44.6\% & 76.6\%\\
Nigeria & 261,895 & 47,770 & 106,339 & 31.0\% & 58.8\%\\
Norway & 121,078 & 41,343 & 52,325 & 44.1\% & 77.4\%\\
Oman & 15,633 & 4,635 & 4,749 & 49.4\% & 60.0\%\\
Papua New Guinea & 2,182 & 525 & 1,009 & 34.2\% & 70.3\%\\
Philippines & 79,309 & 30,165 & 26,658 & 53.1\% & 71.6\%\\
Poland & 359,952 & 129,861 & 121,841 & 51.6\% & 69.9\%\\
Portugal & 149,408 & 67,802 & 58,597 & 53.6\% & 84.6\%\\
Romania & 117,441 & 49,150 & 39,453 & 55.5\% & 75.4\%\\
Russian Federation & 940,103 & 180,330 & 257,350 & 41.2\% & 46.6\%\\
Saudi Arabia & 157,392 & 49,126 & 66,259 & 42.6\% & 73.3\%\\
Serbia & 57,701 & 25,065 & 22,165 & 53.1\% & 81.9\%\\
Somalia & 1,057 & 101 & 325 & 23.7\% & 40.3\%\\
South Africa & 141,129 & 45,024 & 55,139 & 45.0\% & 71.0\%\\
Spain & 801,513 & 291,516 & 310,020 & 48.5\% & 75.1\%\\
Sri Lanka & 43,124 & 4,210 & 6,172 & 40.6\% & 24.1\%\\
Sweden & 207,170 & 69,994 & 84,672 & 45.3\% & 74.7\%\\
Switzerland & 245,044 & 70,646 & 131,979 & 34.9\% & 82.7\%\\
Syrian Arab Republic & 8,132 & 2,525 & 3,694 & 40.6\% & 76.5\%\\
Tunisia & 62,780 & 22,157 & 18,328 & 54.7\% & 64.5\%\\
T\"{u}rkiye & 440,360 & 183,318 & 208,001 & 46.8\% & 88.9\%\\
United Arab Emirates & 34,293 & 9,507 & 10,544 & 47.4\% & 58.5\%\\
United Kingdom & 1,282,607 & 389,777 & 490,191 & 44.3\% & 68.6\%\\
United States & 6,560,519 & 2,119,214 & 2,711,315 & 43.9\% & 73.6\%\\
Zimbabwe & 10,497 & 2,495 & 4,781 & 34.3\% & 69.3\%\\
\hline
\end{longtable}

\newpage

\begin{table}[h]
\centering
\caption{Estimated slope ($\beta_c$) and coefficient of determination ($R^2$) for each country. Standard errors (SEs) and $p$-values are computed using a Newey-West estimator to account for potential autocorrelation.}
\label{table:s8}
\begin{tabular}{l c c c c}
\hline
Country & $\beta_c$ & $R^2$ & SE & $p$-value \\
\hline
Algeria & 0.1537 & 0.9726 & 0.0082 & < 0.0001 \\
Egypt & 0.1473 & 0.9938 & 0.0031 & < 0.0001 \\
Iraq & 0.2613 & 0.9769 & 0.0099 & < 0.0001 \\
Jordan & 0.1429 & 0.9933 & 0.0021 & < 0.0001 \\
Lebanon & 0.1210 & 0.9829 & 0.0053 & < 0.0001 \\
Morocco & 0.1063 & 0.9874 & 0.0032 & < 0.0001 \\
Saudi Arabia & 0.1771 & 0.9800 & 0.0067 & < 0.0001 \\
Tunisia & 0.1262 & 0.9253 & 0.0115 & < 0.0001 \\
United Arab Emirates & 0.1569 & 0.9973 & 0.0015 & < 0.0001 \\
\hline
Australia & 0.0572 & 0.9564 & 0.0040 & < 0.0001 \\
Canada & 0.0502 & 0.9324 & 0.0045 & < 0.0001 \\
France & 0.0418 & 0.9207 & 0.0039 & < 0.0001 \\
Germany & 0.0519 & 0.9549 & 0.0036 & < 0.0001 \\
Italy & 0.0576 & 0.9754 & 0.0030 & < 0.0001 \\
Japan & 0.0192 & 0.9042 & 0.0022 & < 0.0001 \\
Spain & 0.0696 & 0.9659 & 0.0041 & < 0.0001 \\
Sweden & 0.0423 & 0.9521 & 0.0032 & < 0.0001 \\
United Kingdom & 0.0473 & 0.9660 & 0.0029 & < 0.0001 \\
United States & 0.0420 & 0.9678 & 0.0026 & < 0.0001 \\
\hline
\end{tabular}
\end{table}

\newpage

\begin{table}[h]
\centering
\caption{Estimated slope ($\gamma_c$) and coefficient of determination ($R^2$) for each country. Standard errors (SEs) and $p$-values are computed using a Newey-West estimator to account for potential autocorrelation.}
\label{table:s9}
\begin{tabular}{l c c c c}
\hline
Country & $\gamma_c$ & $R^2$ & SE & $p$-value \\ \hline
Algeria & 0.0075 & 0.9598 & 0.0004 & < 0.0001 \\
Egypt & 0.0086 & 0.9719 & 0.0003 & < 0.0001 \\
Iraq & 0.0082 & 0.8495 & 0.0007 & < 0.0001 \\
Jordan & 0.0110 & 0.9487 & 0.0006 & < 0.0001 \\
Lebanon & 0.0075 & 0.9048 & 0.0007 & < 0.0001 \\
Morocco & 0.0029 & 0.6891 & 0.0005 & < 0.0001 \\
Saudi Arabia & 0.0106 & 0.9138 & 0.0010 & < 0.0001 \\
Tunisia & 0.0139 & 0.9759 & 0.0006 & < 0.0001 \\
United Arab Emirates & 0.0098 & 0.9700 & 0.0003 & < 0.0001 \\
\hline
Australia & 0.0069 & 0.9833 & 0.0003 & < 0.0001 \\
Canada & 0.0071 & 0.9951 & 0.0001 & < 0.0001 \\
France & 0.0061 & 0.9716 & 0.0003 & < 0.0001 \\
Germany & 0.0074 & 0.9861 & 0.0003 & < 0.0001 \\
Italy & 0.0048 & 0.9604 & 0.0003 & < 0.0001 \\
Japan & 0.0041 & 0.9940 & 0.0001 & < 0.0001 \\
Spain & 0.0065 & 0.9849 & 0.0002 & < 0.0001 \\
Sweden & 0.0060 & 0.9800 & 0.0002 & < 0.0001 \\
United Kingdom & 0.0066 & 0.9970 & 0.0001 & < 0.0001 \\
United States & 0.0072 & 0.9956 & 0.0001 & < 0.0001 \\
\hline
\end{tabular}
\end{table}

\newpage

\begin{longtable}{l rr rr rr}
\caption{Inter-publication interval thresholds (in years) for female and male authors when the survival probability threshold is set to 1\%, 2\%, and 5\%.} \\
\label{table:s10} \\
\hline
 & \multicolumn{2}{c}{1\%} & \multicolumn{2}{c}{2\%} & \multicolumn{2}{c}{5\%} \\
\cmidrule(lr){2-3} \cmidrule(lr){4-5} \cmidrule(lr){6-7}
Country & \multicolumn{1}{c}{Female} & \multicolumn{1}{c}{Male} & \multicolumn{1}{c}{Female} & \multicolumn{1}{c}{Male} & \multicolumn{1}{c}{Female} & \multicolumn{1}{c}{Male} \\
\hline\endfirsthead
Albania & 8.18 & 9.88 & 6.92 & 7.09 & 4.67 & 4.33 \\
Algeria & 10.12 & 9.50 & 7.75 & 6.68 & 5.01 & 3.98 \\
Australia & 11.93 & 9.26 & 6.88 & 5.26 & 3.08 & 2.36 \\
Austria & 10.40 & 7.85 & 6.16 & 4.52 & 3.02 & 2.09 \\
Bahrain & 13.93 & 13.80 & 9.76 & 8.53 & 5.09 & 4.01 \\
Belarus & 11.52 & 11.27 & 8.46 & 7.45 & 4.70 & 3.83 \\
Belgium & 7.59 & 5.83 & 4.40 & 3.25 & 2.09 & 1.58 \\
Bosnia and Herzegovina & 10.01 & 10.01 & 7.01 & 6.93 & 4.05 & 4.01 \\
Bulgaria & 12.01 & 13.64 & 7.93 & 8.56 & 4.01 & 4.07 \\
Canada & 13.76 & 10.51 & 8.20 & 6.17 & 3.97 & 2.92 \\
Cyprus & 8.76 & 6.26 & 5.79 & 4.09 & 3.22 & 2.16 \\
Czechia & 8.01 & 7.92 & 5.16 & 4.68 & 2.84 & 2.25 \\
Denmark & 10.61 & 7.92 & 6.10 & 4.50 & 3.00 & 2.01 \\
Egypt & 13.75 & 14.01 & 8.68 & 8.67 & 4.61 & 4.21 \\
Estonia & 8.72 & 7.74 & 5.95 & 5.01 & 3.43 & 2.68 \\
France & 10.08 & 8.92 & 6.13 & 5.25 & 3.17 & 2.59 \\
Germany & 12.68 & 9.01 & 7.17 & 5.07 & 3.36 & 2.33 \\
Ghana & 10.47 & 9.01 & 6.89 & 6.01 & 4.01 & 3.17 \\
Iceland & 10.01 & 9.26 & 6.82 & 5.92 & 3.73 & 3.01 \\
Iran, Islamic Republic of & 8.18 & 6.92 & 6.01 & 4.72 & 3.57 & 2.46 \\
Iraq & 10.93 & 11.60 & 7.87 & 7.74 & 4.73 & 4.17 \\
Ireland & 15.65 & 11.93 & 9.01 & 6.67 & 4.01 & 2.91 \\
Israel & 12.01 & 9.18 & 7.36 & 5.47 & 3.81 & 2.67 \\
Italy & 9.18 & 7.51 & 5.17 & 4.22 & 2.45 & 2.00 \\
Jamaica & 20.03 & 20.10 & 12.60 & 13.35 & 6.19 & 6.59 \\
Japan & 16.58 & 10.25 & 10.28 & 6.01 & 4.52 & 2.50 \\
Jordan & 8.75 & 9.08 & 6.09 & 6.24 & 3.54 & 3.42 \\
Kenya & 10.06 & 10.60 & 7.25 & 7.01 & 4.15 & 4.00 \\
Kuwait & 12.04 & 11.35 & 8.81 & 7.50 & 4.89 & 4.01 \\
Lebanon & 11.93 & 11.01 & 7.09 & 6.84 & 3.61 & 3.17 \\
Lithuania & 6.67 & 7.09 & 4.92 & 4.98 & 3.00 & 2.77 \\
Moldova, Republic of & 10.09 & 10.43 & 7.01 & 7.01 & 4.01 & 4.01 \\
Montenegro & 10.52 & 10.01 & 7.01 & 6.84 & 4.01 & 3.59 \\
Morocco & 8.77 & 8.71 & 5.71 & 5.62 & 3.01 & 2.87 \\
Netherlands & 8.56 & 7.82 & 5.16 & 4.64 & 2.67 & 2.17 \\
New Zealand & 15.01 & 10.57 & 8.59 & 6.18 & 4.00 & 2.92 \\
Nigeria & 10.38 & 10.30 & 7.05 & 6.98 & 4.05 & 3.85 \\
Norway & 10.01 & 9.01 & 6.32 & 5.56 & 3.30 & 2.76 \\
Oman & 11.01 & 8.73 & 8.10 & 6.13 & 4.64 & 3.18 \\
Philippines & 14.36 & 13.10 & 9.87 & 8.53 & 5.62 & 4.42 \\
Poland & 9.01 & 8.39 & 5.67 & 5.11 & 3.01 & 2.83 \\
Portugal & 9.92 & 9.01 & 6.02 & 5.42 & 3.01 & 2.52 \\
Romania & 9.77 & 9.01 & 5.90 & 5.51 & 2.94 & 2.82 \\
Russian Federation & 10.01 & 11.30 & 6.47 & 6.92 & 3.34 & 3.25 \\
Saudi Arabia & 8.45 & 7.95 & 5.69 & 5.05 & 3.13 & 2.59 \\
Serbia & 9.51 & 9.84 & 5.84 & 6.01 & 3.00 & 3.01 \\
South Africa & 13.25 & 11.34 & 7.92 & 6.93 & 3.90 & 3.26 \\
Spain & 10.89 & 9.10 & 6.73 & 5.50 & 3.12 & 2.58 \\
Sri Lanka & 10.09 & 12.01 & 7.09 & 8.26 & 3.98 & 4.26 \\
Sweden & 10.01 & 7.76 & 6.05 & 4.64 & 3.06 & 2.18 \\
Switzerland & 11.31 & 8.79 & 6.68 & 5.01 & 3.21 & 2.34 \\
Syrian Arab Republic & 16.56 & 16.01 & 11.09 & 10.66 & 5.34 & 5.64 \\
Tunisia & 8.26 & 7.68 & 5.71 & 5.01 & 3.23 & 2.60 \\
T\"{u}rkiye & 7.34 & 7.02 & 5.01 & 4.73 & 2.77 & 2.50 \\
United Arab Emirates & 10.75 & 8.01 & 6.92 & 5.13 & 3.67 & 2.58 \\
United Kingdom & 15.52 & 11.01 & 8.96 & 6.25 & 3.96 & 2.76 \\
United States & 14.27 & 11.84 & 8.51 & 7.01 & 4.01 & 3.09 \\
Zimbabwe & 12.01 & 10.76 & 8.11 & 7.44 & 4.84 & 4.22 \\
\hline
\end{longtable}

\newpage
\begin{longtable}{l cc cc}
\caption{Spearman's rank correlation between cumulative productivity and two conventional career metrics, total productivity (TP) and publishing-career length (PCL). } \\
\label{table:s11} \\
\toprule
& \multicolumn{2}{c}{Female} & \multicolumn{2}{c}{Male} \\
\cmidrule(lr){2-3} \cmidrule(lr){4-5}
Country & TP & PCL & TP & PCL \\
\hline\endfirsthead
\hline
Country & TP & PCL & TP & PCL \\
\hline\endhead
Albania & 0.961 & 0.689 & 0.965 & 0.718 \\
Algeria & 0.975 & 0.636 & 0.973 & 0.698 \\
Australia & 0.957 & 0.677 & 0.937 & 0.662 \\
Austria & 0.951 & 0.672 & 0.924 & 0.657 \\
Bahrain & 0.986 & 0.504 & 0.969 & 0.533 \\
Belarus & 0.977 & 0.540 & 0.957 & 0.573 \\
Belgium & 0.908 & 0.648 & 0.882 & 0.619 \\
Bosnia and Herzegovina & 0.972 & 0.700 & 0.974 & 0.720 \\
Bulgaria & 0.976 & 0.677 & 0.971 & 0.663 \\
Canada & 0.958 & 0.644 & 0.934 & 0.639 \\
Cyprus & 0.970 & 0.700 & 0.947 & 0.730 \\
Czechia & 0.947 & 0.693 & 0.949 & 0.694 \\
Denmark & 0.957 & 0.699 & 0.933 & 0.677 \\
Egypt & 0.982 & 0.619 & 0.976 & 0.631 \\
Estonia & 0.957 & 0.723 & 0.944 & 0.687 \\
France & 0.949 & 0.688 & 0.932 & 0.670 \\
Germany & 0.957 & 0.643 & 0.924 & 0.639 \\
Ghana & 0.978 & 0.604 & 0.975 & 0.677 \\
Iceland & 0.957 & 0.698 & 0.952 & 0.719 \\
Iran, Islamic Republic of & 0.961 & 0.648 & 0.944 & 0.706 \\
Iraq & 0.981 & 0.539 & 0.979 & 0.571 \\
Ireland & 0.970 & 0.651 & 0.948 & 0.668 \\
Israel & 0.954 & 0.659 & 0.938 & 0.679 \\
Italy & 0.942 & 0.656 & 0.933 & 0.648 \\
Jamaica & 0.977 & 0.611 & 0.966 & 0.613 \\
Japan & 0.945 & 0.563 & 0.892 & 0.556 \\
Jordan & 0.973 & 0.566 & 0.963 & 0.673 \\
Kenya & 0.975 & 0.636 & 0.969 & 0.673 \\
Kuwait & 0.970 & 0.602 & 0.955 & 0.661 \\
Lebanon & 0.973 & 0.616 & 0.966 & 0.672 \\
Lithuania & 0.954 & 0.754 & 0.945 & 0.732 \\
Moldova, Republic of & 0.978 & 0.614 & 0.977 & 0.638 \\
Montenegro & 0.978 & 0.714 & 0.978 & 0.710 \\
Morocco & 0.969 & 0.616 & 0.971 & 0.678 \\
Netherlands & 0.945 & 0.693 & 0.921 & 0.669 \\
New Zealand & 0.967 & 0.674 & 0.948 & 0.680 \\
Nigeria & 0.975 & 0.552 & 0.973 & 0.608 \\
Norway & 0.960 & 0.715 & 0.944 & 0.694 \\
Oman & 0.973 & 0.485 & 0.949 & 0.652 \\
Philippines & 0.982 & 0.473 & 0.979 & 0.504 \\
Poland & 0.966 & 0.723 & 0.957 & 0.703 \\
Portugal & 0.965 & 0.705 & 0.951 & 0.687 \\
Romania & 0.965 & 0.710 & 0.960 & 0.716 \\
Russian Federation & 0.981 & 0.632 & 0.976 & 0.618 \\
Saudi Arabia & 0.979 & 0.567 & 0.966 & 0.662 \\
Serbia & 0.974 & 0.738 & 0.973 & 0.729 \\
South Africa & 0.971 & 0.676 & 0.960 & 0.675 \\
Spain & 0.954 & 0.666 & 0.939 & 0.661 \\
Sri Lanka & 0.971 & 0.598 & 0.977 & 0.607 \\
Sweden & 0.943 & 0.673 & 0.921 & 0.638 \\
Switzerland & 0.958 & 0.671 & 0.925 & 0.657 \\
Syrian Arab Republic & 0.986 & 0.439 & 0.978 & 0.484 \\
Tunisia & 0.954 & 0.666 & 0.945 & 0.698 \\
T\"{u}rkiye & 0.961 & 0.721 & 0.949 & 0.725 \\
United Arab Emirates & 0.980 & 0.546 & 0.967 & 0.691 \\
United Kingdom & 0.964 & 0.646 & 0.938 & 0.639 \\
United States & 0.959 & 0.624 & 0.937 & 0.614 \\
Zimbabwe & 0.970 & 0.624 & 0.961 & 0.654 \\
\bottomrule
\end{longtable}

\newpage
\begin{longtable}{l c c}
\caption{Proportion of active authors in 2023 with a publication gap experience.} \\
\label{table:s12} \\
\toprule
Country & Female (\%) & Male (\%) \\
\hline\endfirsthead
\hline
Country & Female (\%) & Male (\%) \\
\hline\endhead
Albania & 7.87 & 9.97 \\
Algeria & 5.52 & 10.69 \\
Australia & 17.58 & 25.50 \\
Austria & 13.60 & 23.52 \\
Bahrain & 3.14 & 6.88 \\
Belarus & 5.03 & 9.77 \\
Belgium & 22.79 & 32.30 \\
Bosnia and Herzegovina & 7.97 & 12.04 \\
Bulgaria & 10.76 & 11.76 \\
Canada & 12.75 & 20.47 \\
Cyprus & 11.96 & 20.70 \\
Czechia & 17.25 & 23.76 \\
Denmark & 15.32 & 25.46 \\
Egypt & 6.09 & 8.51 \\
Estonia & 14.11 & 19.89 \\
France & 16.07 & 21.93 \\
Germany & 11.66 & 22.43 \\
Ghana & 4.41 & 8.64 \\
Iceland & 17.09 & 21.55 \\
Iran, Islamic Republic of & 9.08 & 19.17 \\
Iraq & 3.90 & 5.51 \\
Ireland & 12.83 & 19.97 \\
Israel & 11.66 & 19.21 \\
Italy & 22.12 & 28.37 \\
Jamaica & 6.31 & 7.62 \\
Japan & 13.04 & 29.59 \\
Jordan & 5.29 & 10.83 \\
Kenya & 6.83 & 9.34 \\
Kuwait & 6.35 & 11.52 \\
Lebanon & 7.12 & 12.55 \\
Lithuania & 16.29 & 19.46 \\
Moldova, Republic of & 6.63 & 8.67 \\
Montenegro & 9.26 & 14.43 \\
Morocco & 8.63 & 14.20 \\
Netherlands & 17.58 & 25.78 \\
New Zealand & 12.36 & 20.90 \\
Nigeria & 4.81 & 6.36 \\
Norway & 14.67 & 21.29 \\
Oman & 3.89 & 12.22 \\
Philippines & 2.85 & 4.08 \\
Poland & 15.56 & 21.78 \\
Portugal & 20.87 & 25.46 \\
Romania & 12.24 & 15.57 \\
Russian Federation & 7.34 & 8.40 \\
Saudi Arabia & 5.34 & 11.49 \\
Serbia & 16.08 & 17.08 \\
South Africa & 9.72 & 13.53 \\
Spain & 17.11 & 24.76 \\
Sri Lanka & 6.59 & 5.60 \\
Sweden & 20.20 & 29.48 \\
Switzerland & 13.23 & 22.33 \\
Syrian Arab Republic & 2.88 & 4.98 \\
Tunisia & 12.61 & 21.59 \\
T\"{u}rkiye & 13.89 & 21.08 \\
United Arab Emirates & 5.10 & 13.53 \\
United Kingdom & 12.43 & 21.75 \\
United States & 12.48 & 19.49 \\
Zimbabwe & 5.55 & 8.66 \\
\bottomrule
\end{longtable}

\newpage
\begin{longtable}{l c c}
\caption{Proportion of returning authors within newly active authors in 2023.} \\
\label{table:s13} \\
\toprule
Country & Female (\%) & Male (\%) \\
\hline\endfirsthead
\hline
Country & Female (\%) & Male (\%) \\
\hline\endhead
Albania & 17.56 & 18.82 \\
Algeria & 7.35 & 11.07 \\
Australia & 16.87 & 24.22 \\
Austria & 12.51 & 25.09 \\
Bahrain & 1.66 & 6.82 \\
Belarus & 5.94 & 10.71 \\
Belgium & 24.22 & 34.90 \\
Bosnia and Herzegovina & 8.45 & 13.62 \\
Bulgaria & 9.76 & 10.83 \\
Canada & 12.86 & 22.26 \\
Cyprus & 13.08 & 20.73 \\
Czechia & 19.34 & 23.93 \\
Denmark & 14.37 & 25.84 \\
Egypt & 4.42 & 7.18 \\
Estonia & 15.22 & 21.03 \\
France & 17.21 & 22.53 \\
Germany & 11.27 & 23.46 \\
Ghana & 4.29 & 8.65 \\
Iceland & 18.56 & 19.40 \\
Iran, Islamic Republic of & 11.85 & 21.75 \\
Iraq & 4.20 & 6.28 \\
Ireland & 11.94 & 19.56 \\
Israel & 11.75 & 18.59 \\
Italy & 20.46 & 25.42 \\
Jamaica & 13.64 & 4.17 \\
Japan & 13.79 & 34.83 \\
Jordan & 3.66 & 9.64 \\
Kenya & 6.23 & 9.35 \\
Kuwait & 7.65 & 14.32 \\
Lebanon & 5.03 & 7.28 \\
Lithuania & 14.71 & 21.62 \\
Moldova, Republic of & 6.99 & 9.66 \\
Montenegro & 8.97 & 9.76 \\
Morocco & 5.56 & 9.28 \\
Netherlands & 16.32 & 25.55 \\
New Zealand & 13.87 & 20.33 \\
Nigeria & 5.22 & 8.06 \\
Norway & 13.33 & 20.88 \\
Oman & 3.51 & 10.84 \\
Philippines & 2.18 & 3.51 \\
Poland & 16.10 & 21.68 \\
Portugal & 22.42 & 27.22 \\
Romania & 12.72 & 15.65 \\
Russian Federation & 7.54 & 7.62 \\
Saudi Arabia & 3.99 & 7.71 \\
Serbia & 16.89 & 20.07 \\
South Africa & 8.56 & 14.20 \\
Spain & 17.27 & 23.33 \\
Sri Lanka & 4.88 & 6.86 \\
Sweden & 22.98 & 34.45 \\
Switzerland & 11.88 & 22.69 \\
Syrian Arab Republic & 3.54 & 3.02 \\
Tunisia & 13.08 & 18.49 \\
T\"{u}rkiye & 12.42 & 20.08 \\
United Arab Emirates & 3.80 & 8.95 \\
United Kingdom & 12.00 & 23.20 \\
United States & 11.94 & 20.65 \\
Zimbabwe & 7.76 & 10.42 \\
\bottomrule
\end{longtable}

\newpage

\begin{longtable}{lcc}
\caption{Researcher inflow and gender gap in cumulative productivity in 2023.} \\
\label{table:s14} \\
\hline
 & Researcher & Gender gap in \\
Country & inflow & cumulative productivity \\
\hline\endfirsthead
\hline
 & Researcher & Gender gap in \\
Country & inflow & cumulative productivity \\
\hline\endhead
Albania & $12.26\%$ & $-31.89\%$ \\
Algeria & $10.30\%$ & $-46.19\%$ \\
Australia & $7.68\%$ & $-35.75\%$ \\
Austria & $9.53\%$ & $-47.00\%$ \\
Bahrain & $13.86\%$ & $-46.77\%$ \\
Belarus & $6.63\%$ & $-35.28\%$ \\
Belgium & $11.46\%$ & $-40.18\%$ \\
Bosnia and Herzegovina & $8.16\%$ & $-32.45\%$ \\
Bulgaria & $7.11\%$ & $-6.30\%$ \\
Canada & $7.45\%$ & $-40.34\%$ \\
Cyprus & $11.02\%$ & $-50.00\%$ \\
Czechia & $8.44\%$ & $-43.01\%$ \\
Denmark & $8.79\%$ & $-47.70\%$ \\
Egypt & $10.70\%$ & $-23.70\%$ \\
Estonia & $8.27\%$ & $-44.60\%$ \\
France & $9.11\%$ & $-34.78\%$ \\
Germany & $8.49\%$ & $-51.63\%$ \\
Ghana & $14.30\%$ & $-42.75\%$ \\
Iceland & $8.11\%$ & $-36.59\%$ \\
Iran, Islamic Republic of & $9.16\%$ & $-54.02\%$ \\
Iraq & $15.33\%$ & $-26.50\%$ \\
Ireland & $7.35\%$ & $-37.79\%$ \\
Israel & $8.85\%$ & $-40.69\%$ \\
Italy & $9.98\%$ & $-29.86\%$ \\
Jamaica & $4.95\%$ & $-14.81\%$ \\
Japan & $6.64\%$ & $-56.26\%$ \\
Jordan & $15.56\%$ & $-47.23\%$ \\
Kenya & $11.89\%$ & $-30.67\%$ \\
Kuwait & $10.43\%$ & $-39.39\%$ \\
Lebanon & $10.99\%$ & $-42.69\%$ \\
Lithuania & $9.99\%$ & $-32.43\%$ \\
Moldova, Republic of & $11.59\%$ & $-20.74\%$ \\
Montenegro & $7.54\%$ & $-35.03\%$ \\
Morocco & $14.08\%$ & $-38.63\%$ \\
Netherlands & $9.78\%$ & $-40.39\%$ \\
New Zealand & $6.90\%$ & $-42.72\%$ \\
Nigeria & $12.72\%$ & $-22.80\%$ \\
Norway & $8.49\%$ & $-38.37\%$ \\
Oman & $13.51\%$ & $-59.85\%$ \\
Philippines & $17.70\%$ & $-29.79\%$ \\
Poland & $8.14\%$ & $-30.25\%$ \\
Portugal & $8.44\%$ & $-23.57\%$ \\
Romania & $9.83\%$ & $-17.79\%$ \\
Russian Federation & $7.80\%$ & $-15.78\%$ \\
Saudi Arabia & $17.63\%$ & $-48.85\%$ \\
Serbia & $6.98\%$ & $-7.94\%$ \\
South Africa & $8.70\%$ & $-31.44\%$ \\
Spain & $8.66\%$ & $-34.10\%$ \\
Sri Lanka & $10.46\%$ & $1.84\%$ \\
Sweden & $7.09\%$ & $-39.90\%$ \\
Switzerland & $9.09\%$ & $-46.99\%$ \\
Syrian Arab Republic & $15.17\%$ & $-33.67\%$ \\
Tunisia & $11.48\%$ & $-46.08\%$ \\
T\"{u}rkiye & $12.07\%$ & $-35.74\%$ \\
United Arab Emirates & $15.68\%$ & $-59.48\%$ \\
United Kingdom & $7.18\%$ & $-40.19\%$ \\
United States & $7.44\%$ & $-37.03\%$ \\
Zimbabwe & $10.73\%$ & $-29.31\%$ \\
\hline
\end{longtable}

\newpage

\renewcommand{\refname}{Supplementary References}

\end{document}